\documentclass[12pt]{iopart}

\usepackage{amsmath,amsfonts,amssymb}
\usepackage{graphicx,psfrag,epsf}
\usepackage{enumerate}
\def\newblock{\ }%
\usepackage[round]{natbib}
\usepackage{xcolor}
\usepackage{url} 

\begin{document}

\title[]{Understanding North Atlantic Climate Instabilities and Complex Interactions using Data Science}

\author{Alka Yadav$^1$, Sourish Das$^2$,  Anirban Chakraborti$^{1}$, and Sudeep Shukla$^{3}$}

\address{$^1$ School of Computational and Integrative Sciences, Jawaharlal Nehru University, New Delhi-110067, India}
\address{$^2$ Chennai Mathematical Institute, Chennai-603103, India}    
\address{$^3$ AI 4 Water LTD, Orpington, UK}
\eads{\mailto{alkayadav.india@gmail.com}, \mailto{sourish@cmi.ac.in}, \mailto{anirban@jnu.ac.in}, \mailto{sudeep@ai4water.org}}

\begin{abstract}
The North Atlantic Oscillation (NAO) index, a measure of sea-level atmospheric pressure variability, holds significant influence over weather patterns in North America and Northern Europe. A negative (positive) NAO value signifies increased cold air outbreaks and storm occurrences (reduced occurrences) in these regions. NAO, a product of multiple climate factors, demonstrates intricate dynamics with sea surface temperature (SST) and sea ice extent (SIE).
In this study, we adopt a data-driven approach to explore the complex interplay between NAO, SST, and SIE, revealing a critical instability rooted in positive feedback loops among these climate variables. Our statistical machine learning methodology examines the impacts of melting Arctic SIE and rising SST on NAO, thereby understanding the weather patterns across the North Atlantic region.
The skewness analysis yields a negative skewness in NAO across various time intervals—daily, weekly, and monthly. This skewness, coupled with NAO's mean zero stationary nature, accentuates system instability. To capture these dynamics, we formulate a Bayesian Granger-causal dynamic linear model, which effectively updates the predictor-dependent variable relationship over time.
The findings underscore an impending critical instability, indicative of more frequent occurrences of intensely cold climates in eastern North America and northern Europe, theory signifies a notable climate shift. By delving into the intricate feedback mechanisms of NAO, SST, and SIE, our study enhances our comprehension of climate variability, fostering a more informed perspective on the imminent climate changes that lie ahead.

\end{abstract}

{\it Keywords:}  Bayesian Dynamic Linear Model, Granger Causality, Kalman Filter
\newpage

\section{Introduction}


The complex dynamics of Earth's climate system, influenced profoundly by human activities since the Industrial Revolution, is characterized by complex interactions and feedback mechanisms. These dynamics often lead to critical instabilities in the climate system, stemming from reinforcing (positive) feedback loops that amplify changes and destabilize the system, as highlighted by \cite{Halloran2020}. In contrast, stabilizing (negative) feedback mechanisms play a crucial role in maintaining the stability of the Earth's climate system, which supports the existence of complex life forms \citep{Kastens2009}. A deeper understanding of these feedback loops, particularly those that exacerbate climate instabilities, is imperative to address the pressing environmental challenges of biodiversity loss, climate change, and ecosystem degradation.

Among the many climate variables, the North Atlantic Oscillation (NAO), sea surface temperature (SST), and Arctic sea ice extent (SIE) are key indicators of climate variability in the North Atlantic region. The NAO, a widely studied atmospheric pressure index, influences weather patterns in eastern North America and northern Europe. Positive phases of the NAO are associated with fewer storms and milder winters, while negative phases bring increased storm activity and colder air outbreaks \citep{Hurrell1995, Hurrell2010, NAO_Definition, Kwok2000}. These interactions are further linked to variations in SST and SIE, forming a complex web of atmospheric and oceanic processes. Studies have revealed that NAO phases impact SST by altering wind patterns, heat flux, and ocean circulation, leading to significant regional climate changes \citep{rogers2004, Hurrell2010}. Simultaneously, changes in SST and SIE influence atmospheric pressure patterns, further modifying the NAO system, and highlighting the interconnected nature of these variables \citep{eayrs2019}.

The presence of a positive feedback loop among SIE, SST, and NAO has been widely acknowledged. Melting Arctic sea ice reduces the surface albedo, increasing heat absorption by the ocean, which in turn raises SST and accelerates further ice loss \citep{Dall2017}. This loop intensifies atmospheric circulation patterns and impacts NAO variability, as discussed in studies by  \cite{Pan2005, Becker1996, Slonosky2001}. The NAO, in turn, influences Arctic sea ice recovery,  \citep{Warner2018}, and SST distributions, further reinforcing this feedback mechanism \citep{Miettinen2011}. This dynamic interplay is also evident on sub-seasonal and decadal timescales, as shown by  \cite{Guokun2021, Delworth2020, Parkinson2000, Bader2011, Kvamsto2004, Kastens2009, Kwok2000}, with significant implications for regional and global climates.

Existing literature provides evidence of these interactions. For instance, the connection between Arctic sea ice and NAO variability has been extensively examined. Researchers have identified statistically significant correlations between autumn Arctic sea ice anomalies and winter NAO phases, which influence weather patterns and climate variability in Europe and North America \citep{Warner2018, Horvath2021}. Several scientific research communities have identified a significant decrease in SIE in coming years, see \cite{Das2018, Cressey2007}. Moreover, studies suggest that NAO-driven SST anomalies contribute to multi-decadal fluctuations in the Atlantic meridional overturning circulation, affecting Arctic warming and tropical storm activity \citep{Delworth2020}. While many of these insights rely on computationally intensive climate models, they often focus on specific Arctic subregions or limited temporal scales, leaving gaps in the understanding of broader North Atlantic dynamics.

Our research aims to address these gaps by employing advanced data-driven statistical methodologies to investigate the presence of a reinforcing feedback loop involving SIE, SST, and NAO. Unlike traditional climate models, which may assume static relationships between variables over time, our approach incorporates dynamic modeling techniques, such as Bayesian Granger-causal dynamic linear models, to account for temporal variations in predictor-dependent relationships \citep{Migon2010, Das2013}.  This approach directly addresses criticisms highlighted by \cite{Kolstad2019} regarding the limitations of static assumptions in conventional statistical models. 
By adopting this dynamic framework, our study provides a more adaptive and comprehensive analysis of the interconnected North Atlantic climate system.

This study seeks to establish the existence of a positive feedback loop among SIE, SST, and NAO, providing new insights into the instabilities of the North Atlantic climate system. By utilizing advanced statistical models, we aim to contribute to a deeper understanding of the mechanisms driving climate variability. Our findings may help inform strategies to mitigate the challenges posed by rapid Arctic warming and its cascading effects on regional and global climates.

\section{Data}\label{sec:dataset}

\subsection{Description}
This study utilizes three distinct datasets: (a) daily mean Arctic Sea Ice Extent (SIE), (b) daily mean Sea Surface Temperature (SST) of north Atlantic basin, and (c) daily mean North Atlantic Oscillation (NAO) index. The NAO and SST datasets are sourced from the National Oceanic and Atmospheric Administration (NOAA) website \citep{NAO_Data_Source, SST_Data_Source}. The SIE dataset is obtained from the National Snow and Ice Data Centre's website \citep{SIE_Data_Source}. These datasets cover a range of time periods: the NAO data is available from 1950, the SIE data from 1979, and the SST data from 1982. The period under consideration spans from January 1982 to September 2019, covering a duration of 38 years.

\subsection{Exploratory Analyses}

This section encompasses the exploratory data analysis conducted on the Arctic Sea Ice Extent (SIE), Sea Surface Temperature (SST), and North Atlantic Oscillation (NAO).

\begin{figure}[ht]
	\centering
\includegraphics[width=0.3\linewidth]{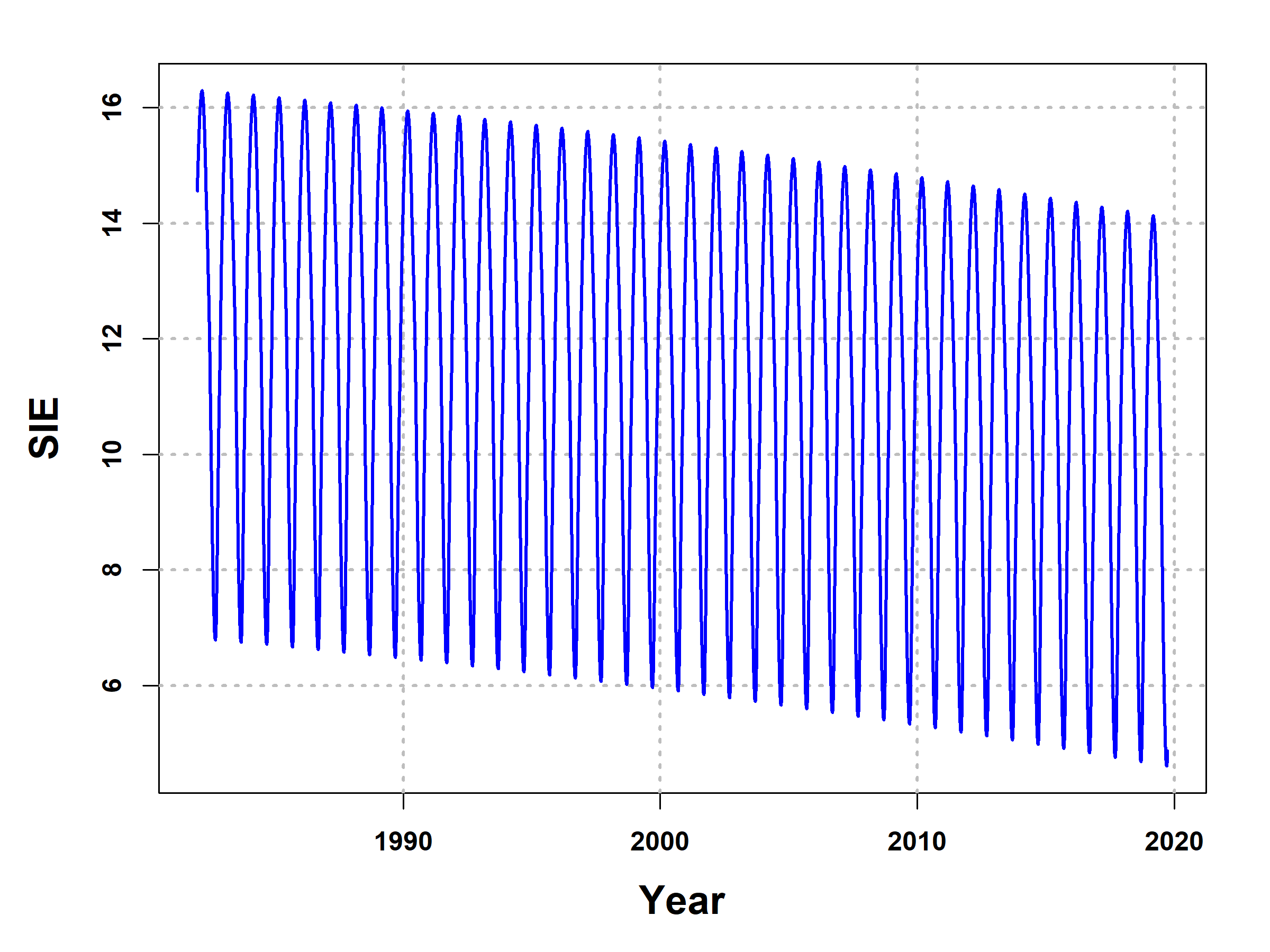}
\llap{\parbox[b]{1in}{\textbf{(a)}\\\rule{0ex}{1.6in}}}
\includegraphics[width=0.3\linewidth]{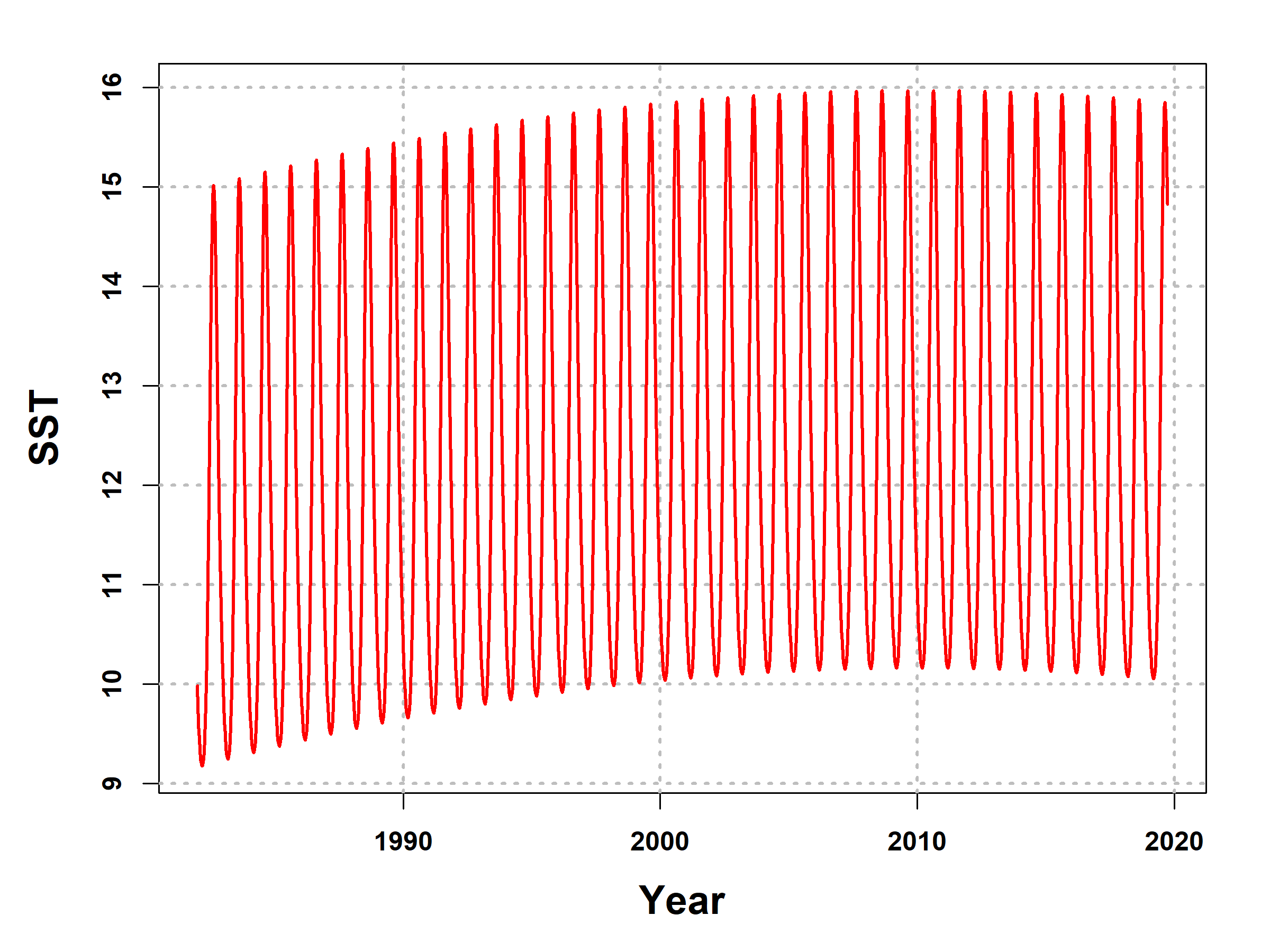}
\llap{\parbox[b]{1in}{\textbf{(b)}\\\rule{0ex}{1.6in}}}
\includegraphics[width=0.3\linewidth]{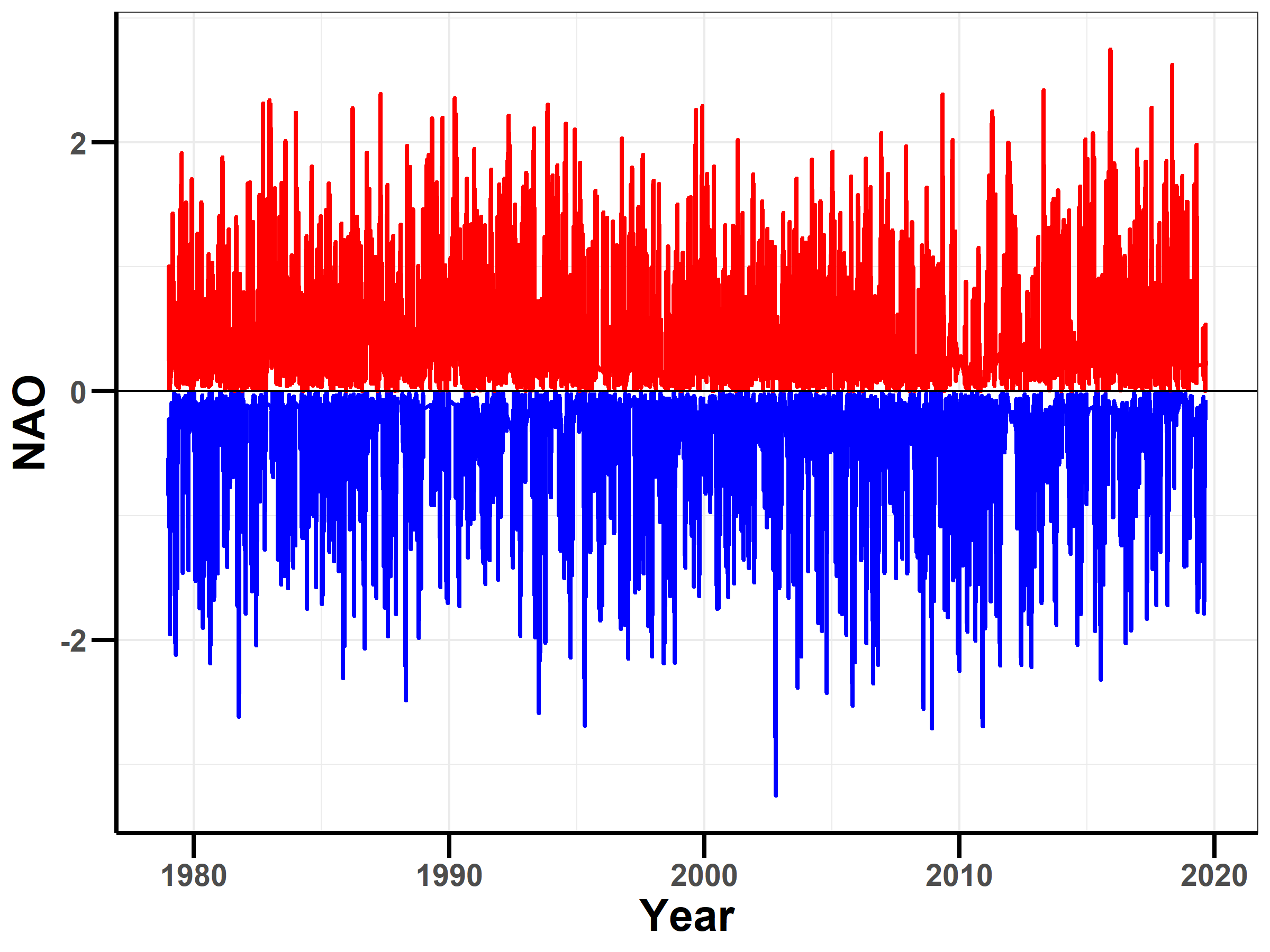}
\llap{\parbox[b]{1in}{\textbf{(c)}\\\rule{0ex}{1.6in}}}
\caption{(a) SIE time series plot from 1982 to 2019. (b) SST time series plot from 1982 to 2019. (c) Plot of the NAO time series over the period of 1979-2019.}
	\label{fig_arctic_seaice}
\end{figure}

In Figure (\ref{fig_arctic_seaice}a), the gradual reduction in SIE from 1982 to 2019 is evident. Notably, during the summer season, Arctic SIE diminishes from 7.412 square km to 3.340 square km. Figure (\ref{fig_arctic_seaice}b) illustrates a consistent upward trajectory in SST over the same timeframe. Complementing this, Figure (\ref{fig_arctic_seaice}c) presents the NAO's time series spanning 1979 to 2019. NAO, representing the difference in sea-level air pressure between the Icelandic Low and the Azores, exhibits a stationary process with a mean of zero.




NAO plays a pivotal role in shaping westerly winds, storm tracks, and climatic conditions across the North Atlantic region. In the positive NAO phase, intensified high and low-pressure systems lead to warmer, wetter winters in northern Europe and northeastern North America. Conversely, the negative NAO phase triggers colder winters and increased Arctic air intrusions in these regions.

\begin{figure}[ht]
    \centering
    \includegraphics[width=0.45\linewidth]{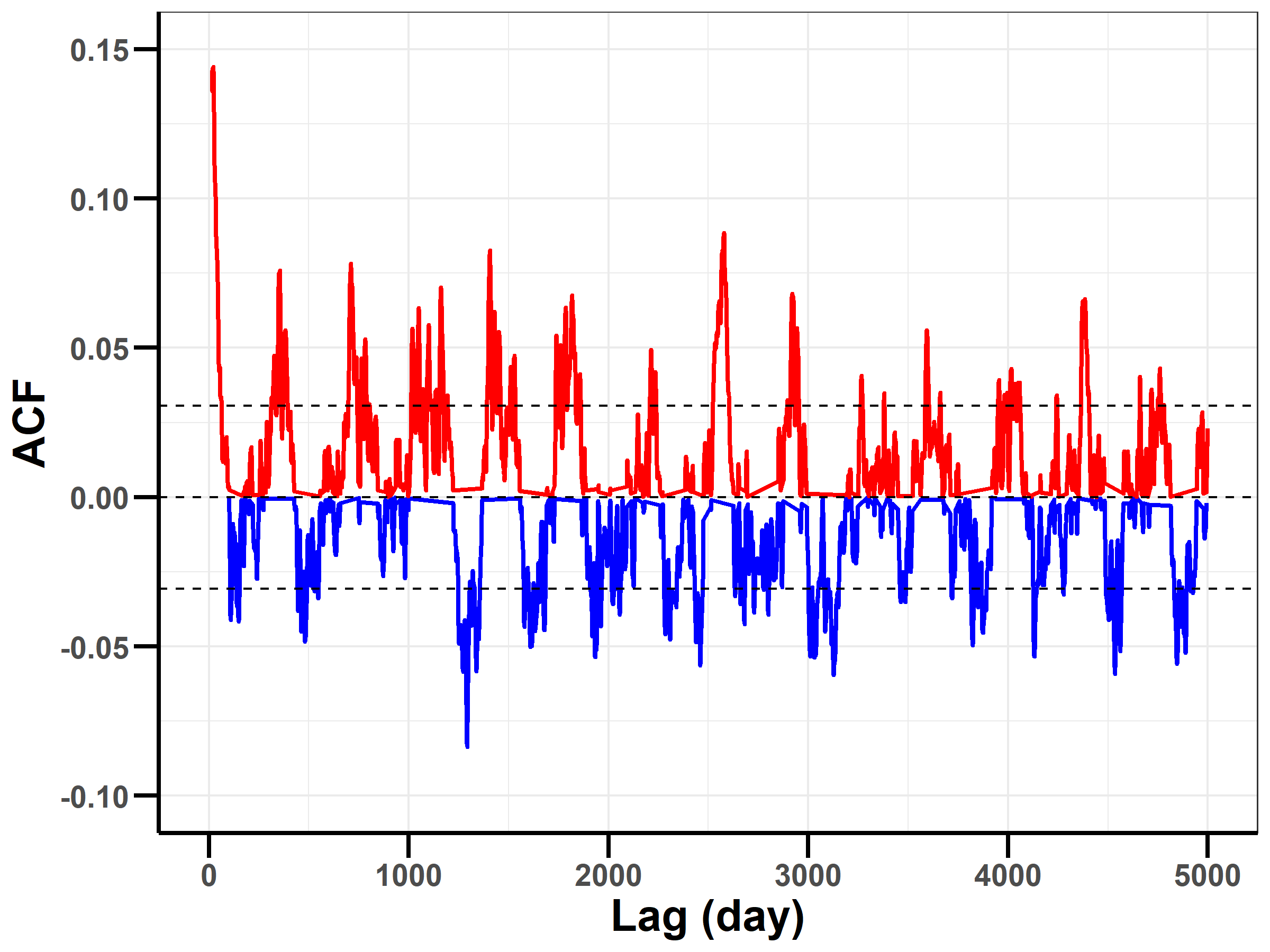}
    \llap{\parbox[b]{1.5in}{\textbf{(a)}\\\rule{0ex}{2.5in}}}
    \includegraphics[width=0.5\linewidth]{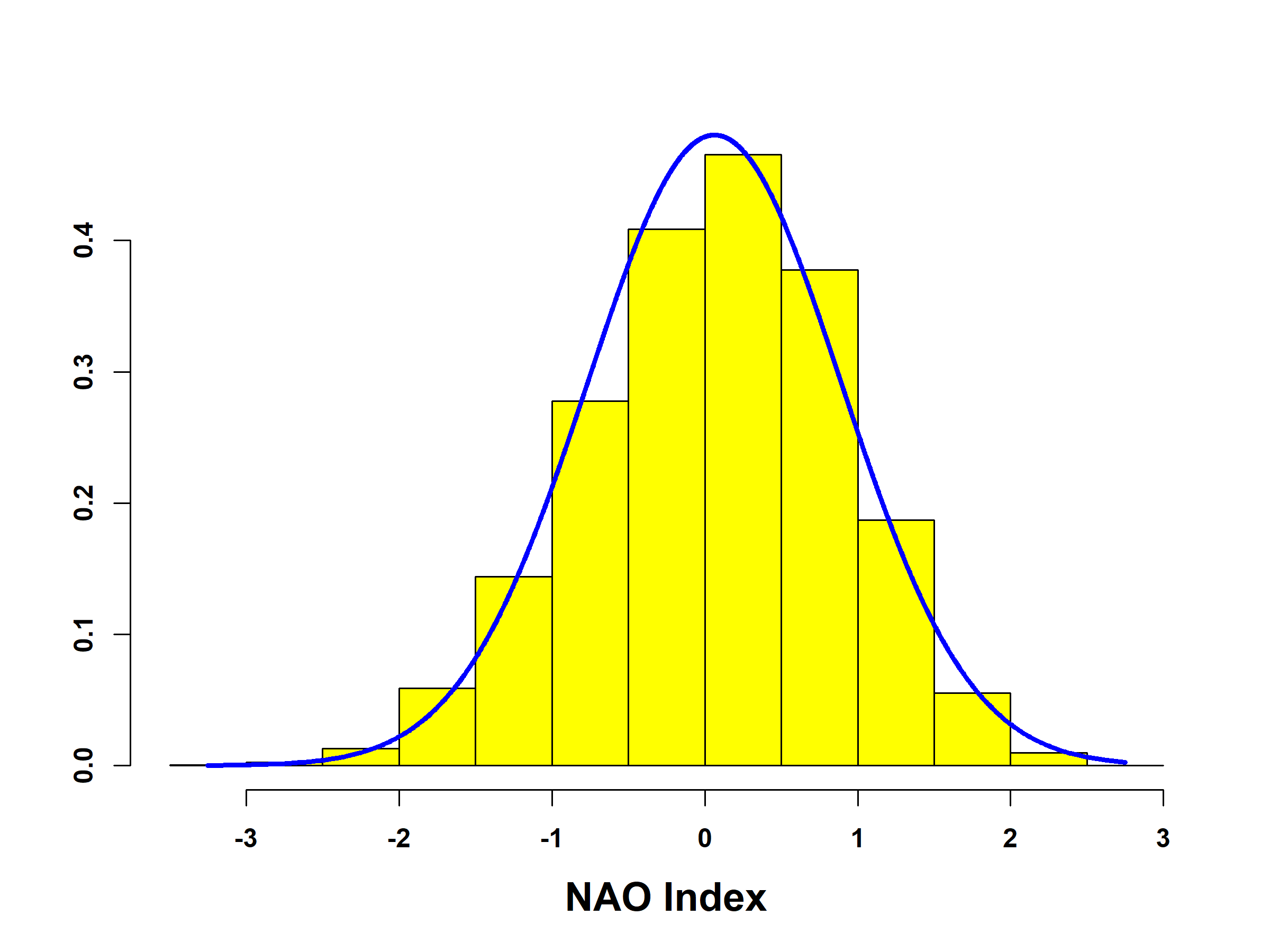}
    \llap{\parbox[b]{1.5in}{\textbf{(b)}\\\rule{0ex}{2.5in}}}
    \caption{ (a) Autocorrelation plot of NAO with a maximum lag of 5000 days (almost 13 years) indicating the existence of long memory.(b) The NAO index's marginal probability distribution from 1979 to 2019 is expected to be bell-shaped and symmetric, with zero skewness, reflecting stability. The empirical evidence indicates instability in the system.}
    \label{fig_Long_memory_of_NAO}
\end{figure}

Figure (\ref{fig_arctic_seaice}c) affirms NAO's mean-zero stationary nature. This was further confirmed through the Augmented Dickey-Fuller test, as indicated by the p-value. Furthermore, the Auto-correlation function (ACF) analysis depicted in Figure (\ref{fig_Long_memory_of_NAO}a) with a maximum lag of 5000 days (approximately 13 years) showing the long memory, and (\ref{fig_Long_memory_of_NAO}b) The figure displays the NAO index's marginal probability distribution, expected to be bell-shaped and symmetric with zero skewness, suggesting stability. However, subsequent empirical evidence, presented later, reveals instability in the system.

Lastly, Table (\ref{tab_Hurst_exp}) showcases the Hurst exponent of the NAO index, significantly surpassing 0.5. This robustly suggests the presence of long memory within the NAO, underscoring its zero-mean stationary characteristics.

\begin{table}[ht]
\begin{center}
	 \setlength{\arrayrulewidth}{.8mm}
	\begin{tabular}{p{8cm}p{2cm}}
	\hline
	\bf{Methods} & \bf{NAO} \\
		\hline
		Simple R/S Hurst estimation  &  0.73\\
		Corrected R over S Hurst exponent  & 0.73\\
		Empirical Hurst exponent  &   0.67 \\
		Corrected empirical Hurst exponent & 0.66 \\
		Theoretical Hurst exponent  & 0.52\\ \hline
	\end{tabular}
		 \caption{The Hurst Exponent value of the NAO index using different methods.}
		 \label{tab_Hurst_exp}
\end{center}
\end{table}

\section{Methodology}\label{sec:methods}

This section outlines the statistical machine learning (SML) methodology employed to elucidate the feedback loop. Figure (\ref{fig_arctic_seaice}a) and Figure (\ref{fig_arctic_seaice}b) illustrate the non-stationary nature of SIE and SST. To address this, we introduce the equations in Section (\ref{SubSec_Modl_SIE}) and (\ref{SubSec_Modl_SST}) that capture the trend and seasonality components of SIE and SST. By de-trending the SIE and SST, we obtain the residuals of both variables. These residuals are expected to exhibit characteristics of a zero-mean stationary process. Consequently, we investigate the Granger-causality between the residuals of SIE, residuals of SST, and the North Atlantic Oscillation (NAO). It is important to note that the NAO was not subjected to the same filtering process as SST to remove seasonal effects because the NAO is inherently a zero-mean stationary process. This characteristic of the NAO indicates the absence of any intrinsic trend or seasonality in the data, making such filtering unnecessary. The Granger-causal model employs auto-regressive time-series techniques to examine the relationships among these variables \citep{Granger1969}.

\subsection{Modeling Sea Ice Extent}\label{SubSec_Modl_SIE}

Let us consider $x(t)$ as the SIE at time point $t$. We formulated the modeling of SIE by incorporating both a trend and a seasonal component. The trend component accounts for long-term variations, indicating either an increase or a decrease over time. On the other hand, the seasonal component captures recurring patterns at a fixed and known frequency, based on specific time periods such as the year, week, or day,
\begin{eqnarray}
 x(t)&=&\underbrace{\beta_0+\beta_1 t + \beta_2 t^2}_{trend} 
 +\underbrace{ \bigg\{\sum_{i=1}^K \alpha_{i}\sin(i\omega t) +\sum_{i=1}^{K}\gamma_{i}\cos(i\omega t)\bigg\}
 }_{seasonality}+ \epsilon(t), 
 \label{eqn_SIE}
 \end{eqnarray}
 
where $\epsilon$ is  error with $\mathbb{E}(\epsilon)=0$, and $Var(\epsilon)=\sigma^2$.  In the seasonality component, we consider the periodicity with $\omega=\frac{2\pi}{365}$; $\alpha_{i}$ is the coefficient corresponding to the sine of $i^{th}$ harmonics and $\gamma_{i}$ is the coefficient corresponding to the cosine of the $i^{th}$ harmonics. Within the model, we take into consideration $K$ harmonics for each period. Now, two questions arise: (i) How do we determine the appropriate value of $K$? and (ii) When $K$ becomes large, several harmonics may become redundant. To address the first question, we fit the model using various values of $K$ ranging from 1 to $K_0$, and select the model with the minimum out-of-sample root mean square error (RMSE). Next, we utilize the least absolute shrinkage and selection operator (LASSO) technique to identify the optimal harmonics in Model (\ref{eqn_SIE}). The LASSO method selectively retains the harmonics that demonstrate a statistically significant impact in minimizing the error \citep{Tibshirani1996}. A similar technique was successfully used to model long-term memory in climate variables \citep{YADAV2023,YADAV2024}.

\subsection{Modeling Sea Surface Temperature}\label{SubSec_Modl_SST}

Let us consider $y(t)$ as the SST at time point $t$. Like SIE, in Equation (\ref{eqn_SIE}), we formulated the modeling of SST by incorporating both a trend and a seasonal component,
\begin{eqnarray}
 y(t)&=&\underbrace{\tilde{\beta}_0+\tilde{\beta}_1 t + \tilde{\beta}_2 t^2}_{trend} 
 +\underbrace{ \bigg\{\sum_{i=1}^K \tilde{\alpha}_{i}\sin(i\omega t) +\sum_{i=1}^{K}\tilde{\gamma}_{i}\cos(i\omega t)\bigg\}
 }_{seasonality}+ \delta(t), 
 \label{eqn_SST}
\end{eqnarray}
where $\delta$ is  error with $\mathbb{E}(\epsilon)=0$, and $Var(\epsilon)=\sigma^2$. In the seasonality component, we consider the periodicity with $\omega=\frac{2\pi}{365}$; the component and parameters of the Model (\ref{eqn_SST}) have same interpretation as  in the Model (\ref{eqn_SIE}). Therefore, we employ the same strategy to fit the model.

Let us assume that $e(t)$ represents the residual obtained from the best fit of Model (\ref{eqn_SIE}). Henceforth, we will refer to $e(t)$ as Sea Ice Extent Residuals (SIER). Similarly, $d(t)$ denotes the residual obtained from the best fit of Model (\ref{eqn_SST}). Moving forward, we will refer to $d(t)$ as Sea Surface Temperature Residuals (SSTR). It is anticipated that both $e(t)$ and $d(t)$ will exhibit characteristics of zero-mean stationary processes, similar to the North Atlantic Oscillation (NAO). Therefore, our objective is to explore the potential presence of a feedback loop between the SIER, SSTR and NAO by employing the Granger causality test.

\subsection{Formulating Feedback Loop with Granger Causality }\label{feedback loop}

In order to examine the evolving causal dynamics between SIER and SSTR, and vice versa, we formulate the following hypothesis utilizing the Granger causal models. The full model incorporates the consideration of one variable being influenced by its own historical memory as well as the lagged values of the other two variables. For instance, the NAO variable is modeled as a function of its own lagged values and the lagged value of both SIER and SSTR, i.e.,
\begin{eqnarray}
\label{eqn_sier_and_sstr_causes_nao}
n(t)&=&\beta_0+\beta_1 n(t-1)+\cdots+\beta_k n(t-k) \nonumber \\
&&~~+\gamma_1 e(t-1)+\cdots+\gamma_k e(t-k) \nonumber \\
&&~~+\delta_1 d(t-1)+\cdots+\delta_k d(t-k)+\epsilon(t).
\end{eqnarray}
Similarly, SIER is modelled as a function of its own lagged values and the lagged value of both NAO and SSTR, i.e.,
\begin{eqnarray}
\label{eqn_nao_and_sstr_causes_sier}
e(t)&=&\beta_0+\beta_1 e(t-1)+\cdots+\beta_k e(t-k) \nonumber \\
&&~~+\gamma_1 n(t-1)+\cdots+\gamma_k n(t-k) \nonumber \\
&&~~+\delta_1 d(t-1)+\cdots+\delta_k d(t-k)+\epsilon(t).
\end{eqnarray}
Finally, SSTR is modelled as a function of its own lagged values and the lagged value of both NAO and SIER, i.e.,
\begin{eqnarray}
\label{eqn_nao_and_sier_causes_sstr}
d(t)&=&\beta_0+\beta_1 d(t-1)+\cdots+\beta_k d(t-k) \nonumber \\
&&~~+\gamma_1 n(t-1)+\cdots+\gamma_k n(t-k) \nonumber \\
&&~~+\delta_1 e(t-1)+\cdots+\delta_k e(t-k)+\epsilon(t).
\end{eqnarray}
The Equation (\ref{eqn_sier_and_sstr_causes_nao}, \ref{eqn_nao_and_sstr_causes_sier}, \ref{eqn_nao_and_sier_causes_sstr}) collectively form the model depicting the feedback loop between NAO, SIER, and SSTR. To examine the presence of this feedback loop, we conduct the Granger causality test, such as the null hypothesis says that all $\gamma_i=0$, i.e.,
\begin{eqnarray}
    H_0: \gamma_1=\cdots=\gamma_k=0~~~\text{and} ~~~~\delta_1=\cdots=\delta_k=0, \label{hypothesis_null}
\end{eqnarray}
to reject the null hypothesis in our alternate hypothesis we have to check if at least one $\gamma_i~~\text{or}~~\delta_i \neq 0$, i.e., 
\begin{eqnarray}
    H_1: \text{at least one } \gamma_i \neq 0~~~\text{or} ~~~~\delta_i \neq 0. \label{hypothesis_alt}
\end{eqnarray}
We run this test for all three models, and if all three tests are rejected, it confirms the existence of the feedback loop.

\subsection{Dynamic Statistical Approach}\label{BDLM}

 \cite{Kolstad2019} has criticized the statistical models for climate research, citing the reason that the relationship between the predictor and dependent variable did not change over time (i.e., regression coefficients remained constant over time). Here, we take into account this criticism by constructing a Bayesian Dynamic Linear Model, along the lines of the work by \cite{Das2013}, where the relationship between a predictor and the dependent variable is updated over time by updating the coefficients. First, we consider the first derivative (momentum) and second derivative (acceleration) of the model presented in the above Equation (\ref{eqn_SIE}) are:
\begin{eqnarray}
 x'(t)&=&\underbrace{\beta_1  + 2\beta_2 t}_{trend} +\underbrace{\omega \bigg\{ \sum_{i=1}^K \alpha_{i}\cos(i\omega t) -\sum_{i=1}^{K}\gamma_{i}\sin(i\omega t)\bigg\}}_{seasonality} + \epsilon_1(t) 
 \label{eqn_first_derivative_of_SIE}
\end{eqnarray}
\begin{eqnarray}
 x''(t)&=& \underbrace{2\beta_2}_{trend} +\underbrace{\omega^2\bigg\{  \sum_{i=1}^K \alpha_{i}\sin(i\omega t) +\sum_{i=1}^{K}\gamma_{i}\cos(i\omega t)\bigg\}}_{seasonality} + \epsilon_2(t), 
 \label{eqn_second_derivative_of_SIE}
\end{eqnarray}
where $\epsilon_1$ and $\epsilon_2$ are noises corresponding to momentum and acceleration signal. Then we developed the Bayesian dynamic linear models (BDLM) for NAO as follows:
\begin{eqnarray}
n_t &=& \beta_{0t}+\beta_{1t}n_{t-1}+\gamma_{1t}x'_{t-1}+\epsilon_{t}\label{obs_eqn_dlm},\\
\beta_{0t} &=& \rho_0
\beta_{0,t-1}+\eta_{1t},\label{system_eqn_dlm_1}\\
\beta_{1t} &=& \rho_1 \beta_{1,t-1}+\eta_{2t},\label{system_eqn_dlm_2},\\
\gamma_{1t} &=& \rho_2 \gamma_{1,t-1}+\eta_{3t},\label{system_eqn_dlm_3}
\end{eqnarray}
where Equation (\ref{obs_eqn_dlm}) is known as observation equation; and Equation (\ref{system_eqn_dlm_1}, \ref{system_eqn_dlm_2},\ref{system_eqn_dlm_3}) are known as the system equations; the $n_t$ is NAO at time point $t$; the $x'_t$ is the momentum of SIE at time point $t$; $\epsilon_t$ is white noise follows $N(0,\sigma_{\epsilon}^2)$; $\eta_{jt},~j=1,2,3$ are white noise associated with system equation follows $N(0,\sigma_{\eta_j}^2)$. Note that $\rho=(\rho_0,\rho_1,\rho_2)$ should be bounded, that is, $|\rho_j|<1$, such that the model would be a stationary process like the NAO index. We know that NAO is a mean stationary process; see Figure (\ref{fig_arctic_seaice}c); hence this restriction of $\rho_j$ is required. We present the Equation (\ref{obs_eqn_dlm}) and Equation (\ref{system_eqn_dlm_1},\ref{system_eqn_dlm_2},\ref{system_eqn_dlm_3}) in the matrix notations as follows:
\begin{eqnarray}
	Y_t&=&X_t\beta_t +\epsilon_{t},~~~~\text{observation ~equation},\nonumber\\
	\beta_t&=& R \beta_{t-1} + \eta_{t},~~~~\text{system~ equation},\label{system_eqn_dlm_matrix}
\end{eqnarray}
where $\epsilon_{t}\sim N(0,\sigma^2)$, and $\eta_t\sim N(0,Q)$. The Bayesian update solution (aka., Kalman filter) for Equation (\ref{system_eqn_dlm_matrix}) can be taken from \citep{Das2013,Singpurwala1983}, as follows
\begin{eqnarray}
K_t&=&R\Sigma_{t}X_t^T(X_t\Sigma_{t}X^T+\sigma^2)^{-1},\nonumber\\
\hat{\beta}_{t+1}&=&R\hat{\beta}_{t}+K_t(Y_t-X_t\hat{\beta}_t),\label{kalman_filter_eqn}\\
\Sigma_{t+1}&=&R\Sigma_tR^T - K_t X_t\Sigma_{t}R^T+Q.\nonumber
\end{eqnarray}
In this study, BDLM corresponds to the Equation (\ref{eqn_SIE}) developed. These models are then updated by the BDLM using a Kalman filter, as presented in Equation (\ref{kalman_filter_eqn}). Through these Kalman updates, $\beta$ coefficients are dynamically updated. The model for SIE is as follows:
\begin{eqnarray}
x(t)&=& \beta_0(t)+\beta_{1}(t)\sin(\omega t) +\gamma_{1}(t)\cos(\omega t)+ \epsilon,\label{eqn_dlm_SIE}
\end{eqnarray}
where $x(t)$ is SIE extent at time point $t$; $\beta_0(t)$, $\beta_1(t)$ and $\gamma_{1}(t)$ are the dynamic coefficients, updated via Kalman update Equation (\ref{kalman_filter_eqn}).
Note that the model does not have any trend part because if there is any trend in the data, that will be captured automatically in the coefficients. In addition, the BDLM was built in the following way:
\begin{eqnarray}
	n(t)&=& \beta_0(t)+\sum_{k=1}^K\beta_{k}(t)n(t-1) +\sum_{k=1}^{K}\gamma_{k}(t)n(t-k)+ \epsilon(t),\label{eqn_dlm_nao}
\end{eqnarray}
where $n(t)$ is NAO and $x'(t-k)$ is the momentum of the SIE at ($t-k$). Here, the Akaike information criterion type model selection process was used to obtain the optimal choice for the model Equation (\ref{eqn_dlm_nao}). It may be noted that if the model's coefficients are static, then the Granger causal model is a special case of Equation (\ref{eqn_dlm_nao}).

\section{Analyses and Results}\label{sec:results}

\subsection{Analyses of SIER, SSTR, and NAO}
\begin{figure}[ht]
 \centering
 \includegraphics[width=0.4\linewidth]{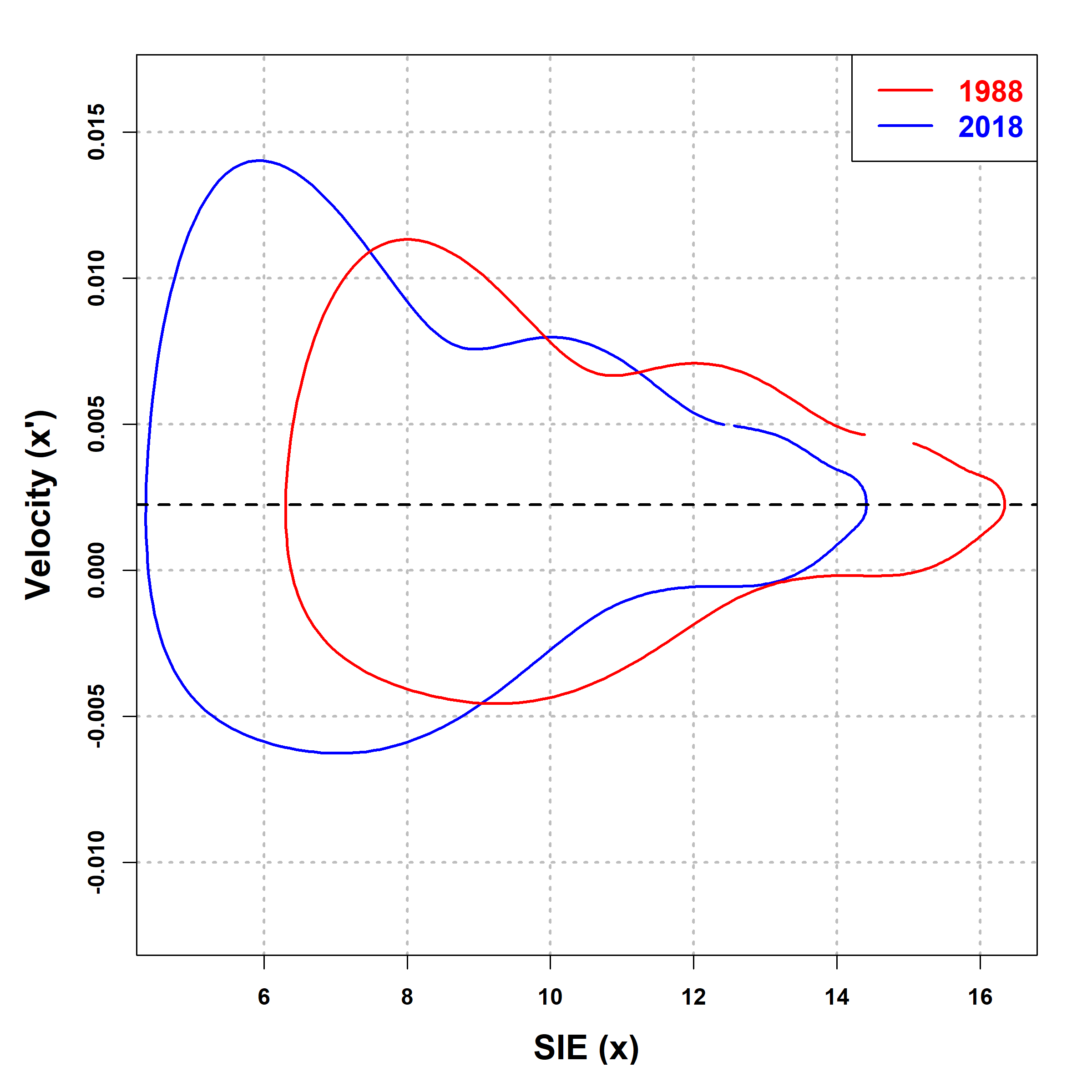}
 \llap{\parbox[b]{2.6in}{\textbf{(a)}\\\rule{0ex}{2.65in}}}
 \includegraphics[width=0.4\linewidth]{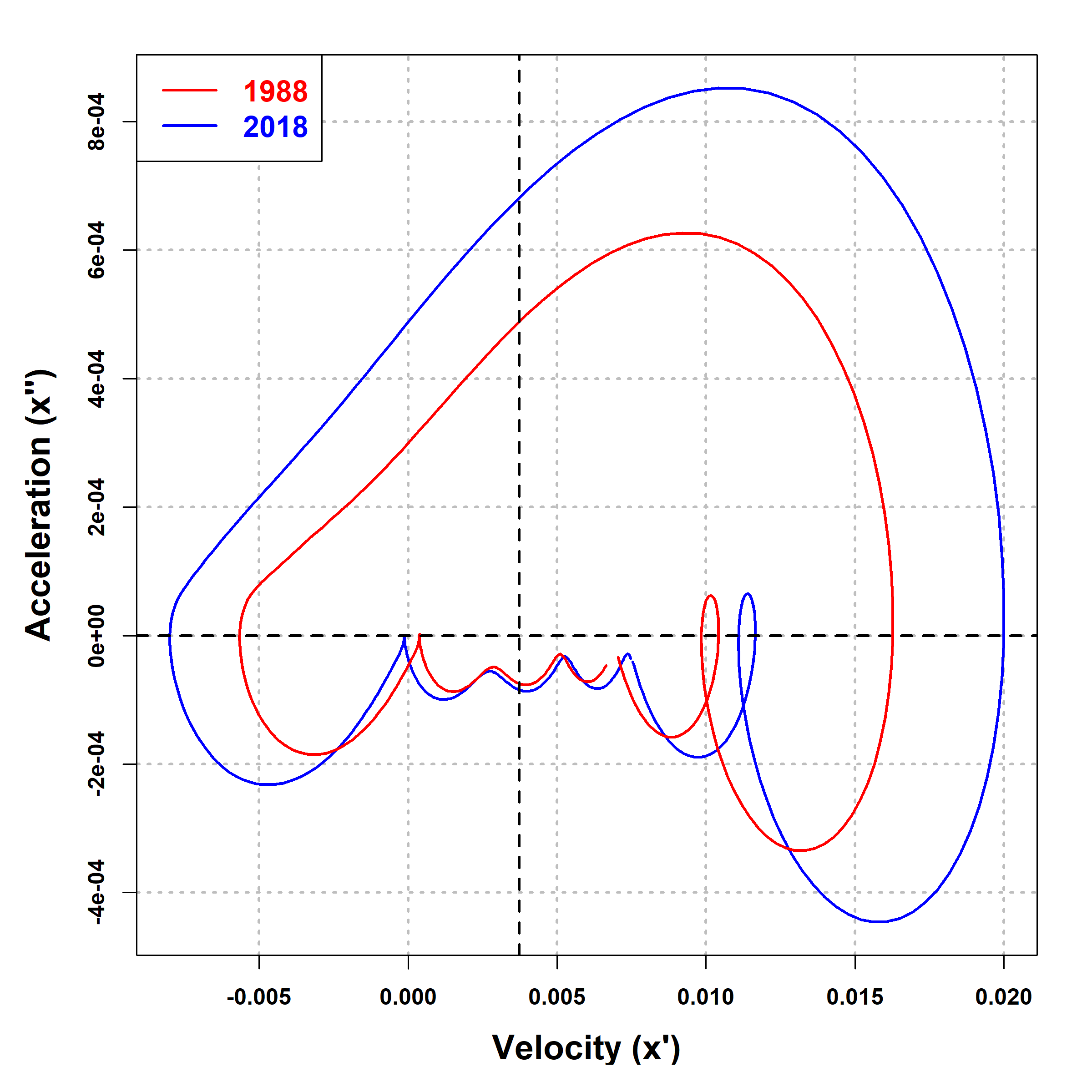}
 \llap{\parbox[b]{2.6in}{\textbf{(b)}\\\rule{0ex}{2.65in}}}
 	\caption{ (a) Phase-line plot of SIE  ($x(t)$ as per Equation (\ref{eqn_SIE}) vs.  velocity of SIE ($x'(t)$ as per Equation (\ref{eqn_first_derivative_of_SIE})  for the two years 1988 and 2018. (b) The phase-plane plot of the velocity of SIE ($x'(t)$ as per Equation (\ref{eqn_first_derivative_of_SIE}) vs. acceleration of SIE ($x''(t)$ as per Equation (\ref{eqn_second_derivative_of_SIE}) for the two years 1988  and 2018. }
 	\label{fig_phase_plane_of_SIE}
\end{figure}

The modeling framework pertaining to SIE and SST is detailed in sections (\ref{SubSec_Modl_SIE}) and (\ref{SubSec_Modl_SST}). In Figure \ref{fig_phase_plane_of_SIE}(a,b), which illustrates the phase-plane analysis of SIE, noticeable fluctuations in both the spatial extent and the rate of volume change of SIE become apparent. These figures portray the dynamic interplay through variations in the phase-line distribution, represented by the expanding area beneath the paired curves. Furthermore, a significant increase in the rate of sea ice melting between 1988 and 2018 becomes evident, an outcome attributed to the phenomenon of melting SIE.

\begin{figure}[h]
   \centering
    \includegraphics[width=0.49\linewidth]{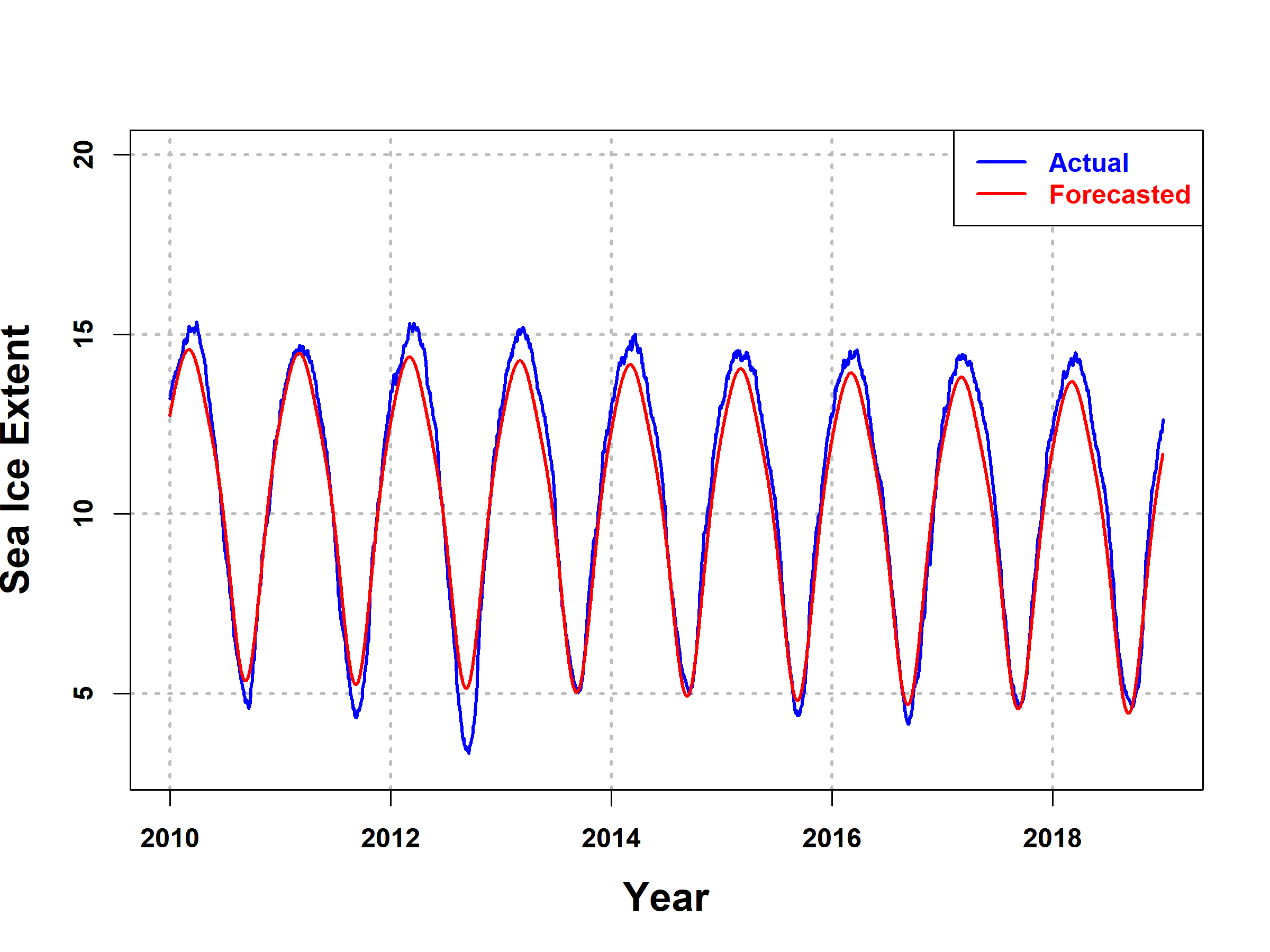}
    \caption{ 
To evaluate the effectiveness of our proposed model (Equations (\ref{eqn_SIE},\ref{eqn_first_derivative_of_SIE},\ref{eqn_second_derivative_of_SIE})), a machine learning assessment was carried out. The training dataset spanned from 1979 to 2009, while the test dataset encompassed the years 2010 to 2019. The R-square value is - 0.985.   }
    \label{test_plot}
\end{figure}

\begin{figure}[ht]
 \centering
 \includegraphics[width=0.4\linewidth]{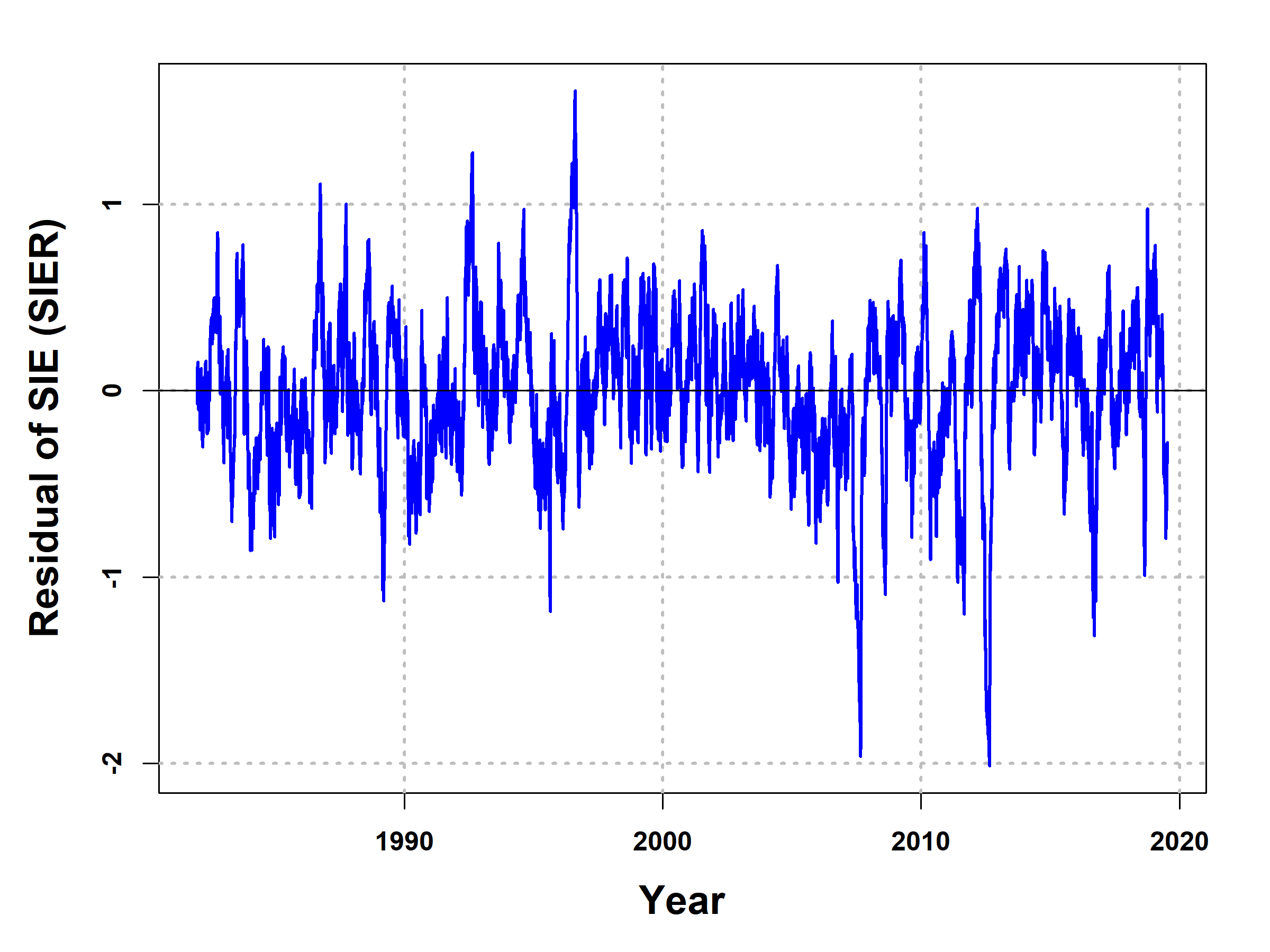}
 \llap{\parbox[b]{2.2in}{\textbf{(a)}\\\rule{0ex}{1.9in}}}
 \includegraphics[width=0.4\linewidth]{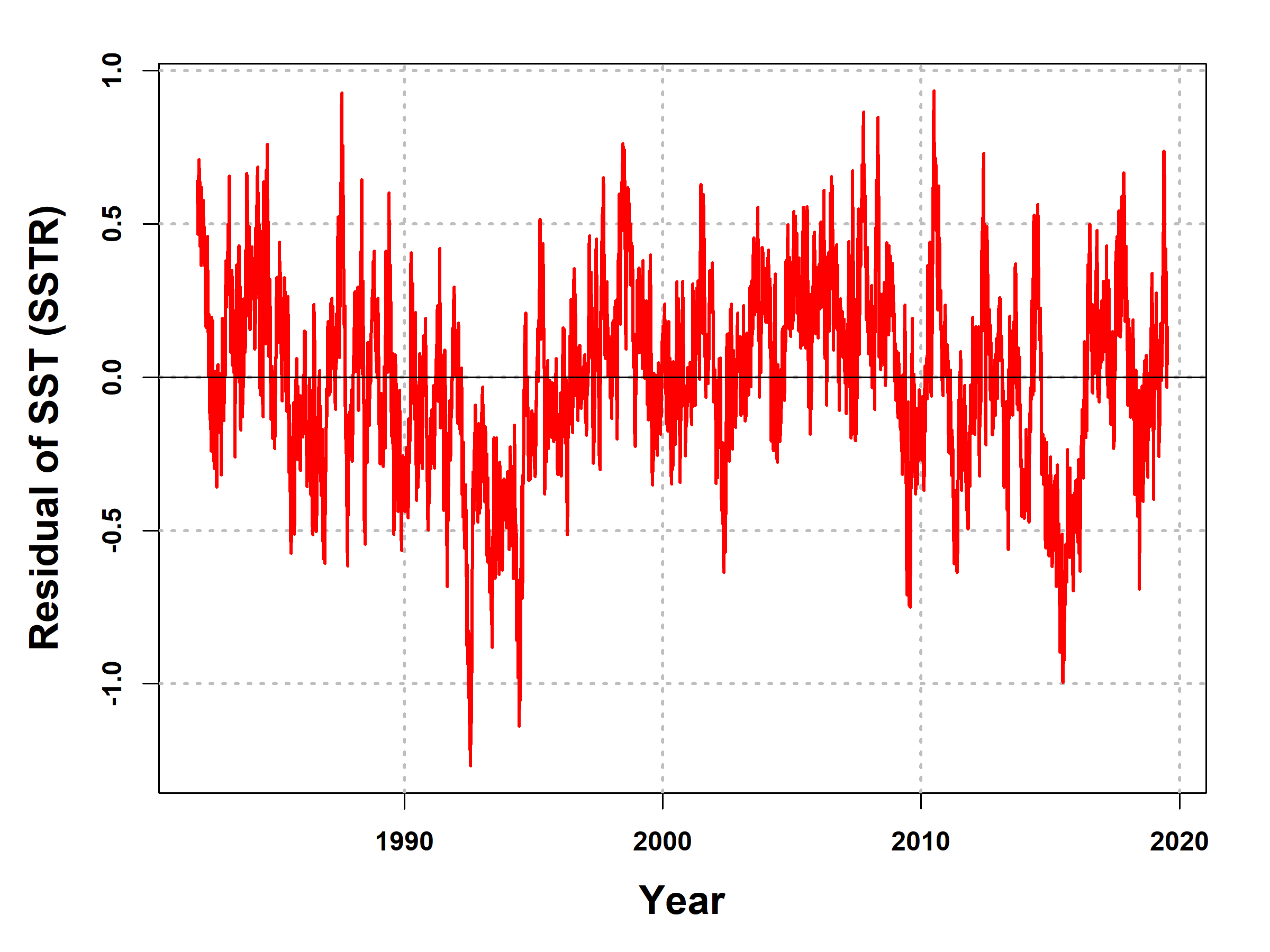}
 \llap{\parbox[b]{2.2in}{\textbf{(b)}\\\rule{0ex}{1.9in}}}
 	\caption{ Time series plots of (a) Residual of SIE (SIER) (b) Residual of SST (SSTR) }
 	\label{fig_ts_of_SIER_SSTR}
\end{figure}

Moving on to Figure (\ref{test_plot}), a visualization of the projected and observed trajectories within the test dataset spanning from 2010 to 2019 is presented. Root Mean Square Error (RMSE) values are computed for both the training and test datasets, yielding an RMSE value of 0.36 for the training set and 0.41 for the test set. This underscores the model's robust capacity for generalization in out-of-sample scenarios, as indicated by the high R-squared value of 0.9865. The time series plot for SIER and SSTR is portrayed in Figure (\ref{fig_ts_of_SIER_SSTR}). Notably, both processes exhibit the characteristic of being mean zero stationary, which is a fundamental requirement for conducting the Granger causal test.

\begin{figure}[ht]
 \centering
 \includegraphics[width=0.4\linewidth]{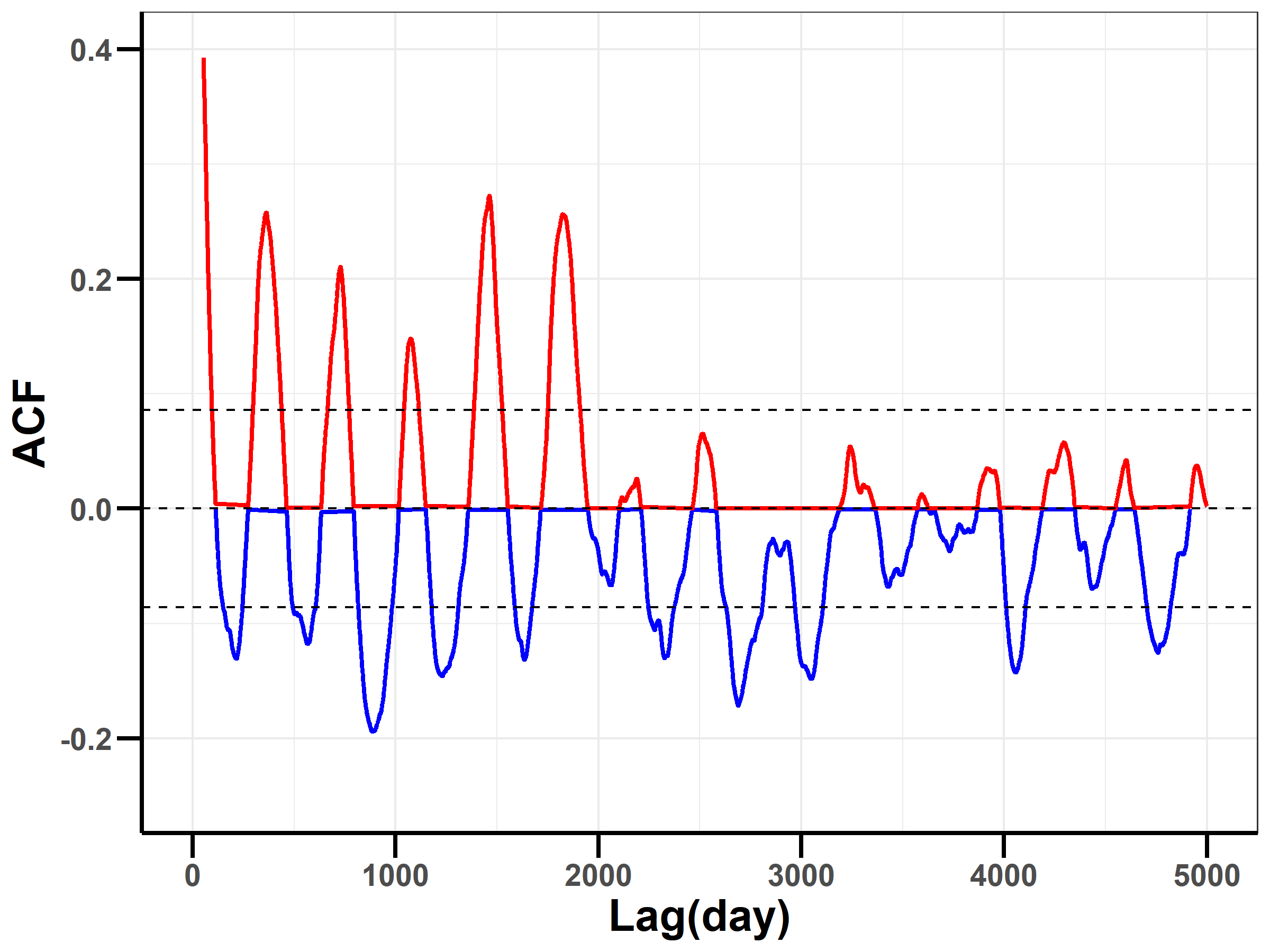}
 \llap{\parbox[b]{2.2in}{\textbf{(a)}\\\rule{0ex}{2in}}}
 \includegraphics[width=0.4\linewidth]{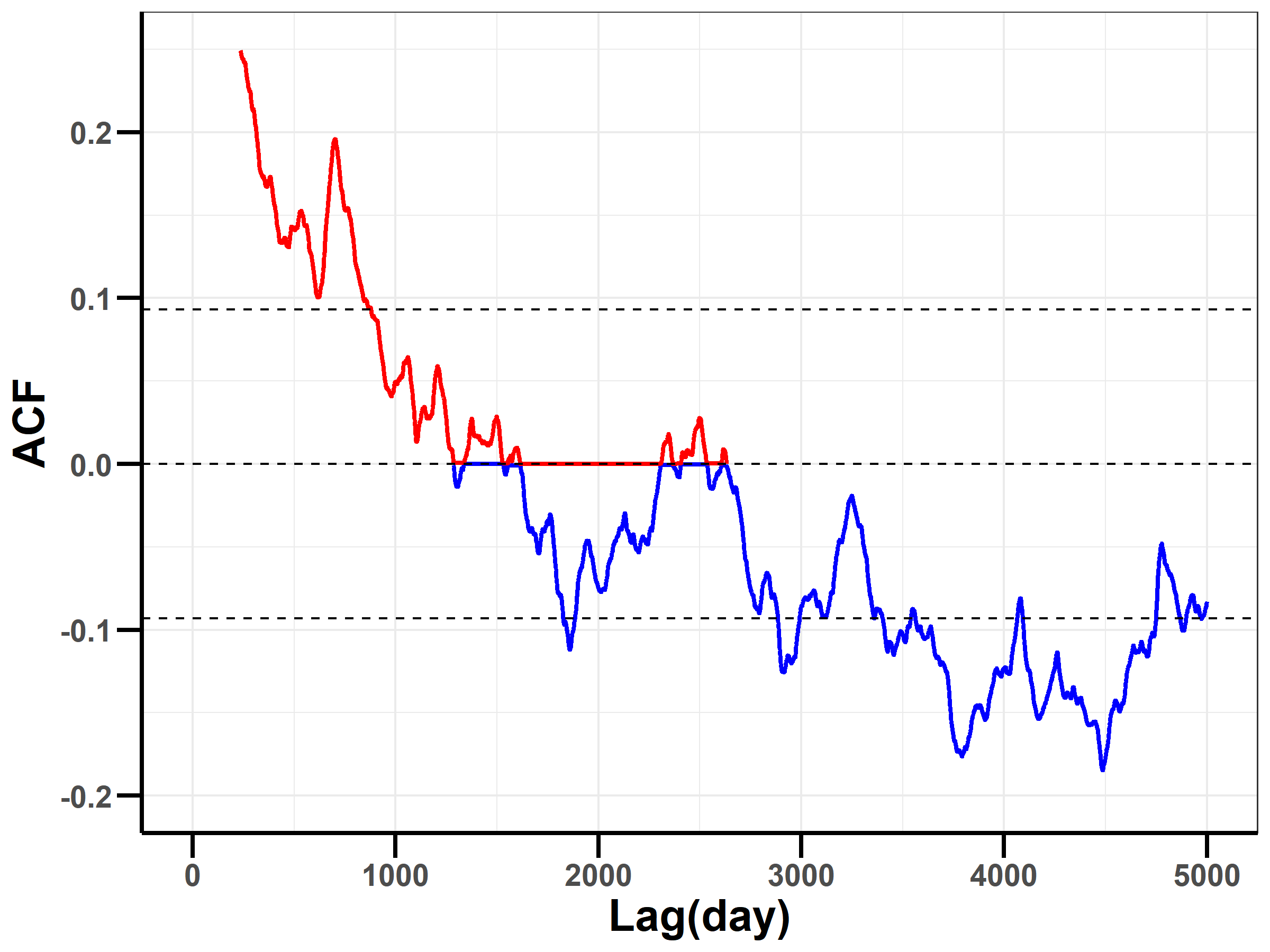}
 \llap{\parbox[b]{2.2in}{\textbf{(b)}\\\rule{0ex}{2in}}}
 	\caption{ ACF plots of (a) Residual of SIE (SIER) (b) Residual of SST (SSTR) }
 	\label{fig_acf_of_SIER_SSTR}
\end{figure}

Figure (\ref{fig_acf_of_SIER_SSTR}) and Table (\ref{tab_Hurst_exp_SIER_SSTR}) jointly demonstrate the presence of long memory in SIER and SSTR. This inference is substantiated by the significantly elevated Hurst exponent values, exceeding the threshold of 0.5, thus reflecting the underlying memory dynamics in these processes. The correlation matrix for NAO, SIER, and SSTR is presented in Table (\ref{tab_NAO_SST_SIE_corr}), calculated over a span of 38 years from January 1982 to September 2019. Enclosed within parentheses are the associated P-values, which offer insights into the statistical significance of these correlations. Particularly noteworthy is the robust and statistically significant correlation between SSTR and NAO. Furthermore, a strong correlation is evident between SIER and SSTR. However, it is important to note that the correlation between NAO and SIER appears to be relatively weaker in significance.

\subsection{Granger Causal Test }
We delve into the examination of the positive feedback loop. 
The Granger causal models are formulated with null and alternative hypotheses, as depicted in Equations (\ref{hypothesis_null}) and (\ref{hypothesis_alt}). The null hypothesis asserts that all $\gamma_i$ coefficients are equal to zero, thereby establishing $H_0: \gamma_1=\cdots=\gamma_k=0$. In contrast, the alternative hypothesis $H_1$ aims to reject this by examining whether at least one $\gamma_i$ deviates from zero. Table (\ref{GC_models_result}) presents the ANOVA F-test outcomes for the Granger causal models outlined in Equations (\ref{eqn_sier_and_sstr_causes_nao}), (\ref{eqn_nao_and_sstr_causes_sier}), and (\ref{eqn_nao_and_sier_causes_sstr}).

The ANOVA F-test ($\text{p-value}= 0.0178$) effectively refutes the null hypothesis, indicating that SSTR and SIER indeed exert a Granger-causal influence on NAO. Similarly, the ANOVA F-test ($\text{p-value}=2.16\times 10^{-6}$) dismisses the notion that NAO and SIER lack a Granger-causal impact on SSTR. Likewise, the ANOVA F-test ($\text{p-value}=2.17\times 10^{-10}$) rejects the null hypothesis that NAO and SSTR do not possess a Granger-causal effect on SIER. These compelling outcomes collectively confirm the existence of a feedback loop connecting SIER, SSTR, and NAO.

Continuing, the analysis we employ an Akaike information criterion-based model selection process, and we identify the optimal configuration for the Model Equation (\ref{eqn_dlm_nao}). Notably, if the model coefficients were static, the Granger causal model would represent a special case of Equation (\ref{eqn_dlm_nao}).

The synthesis of these revelations underscores the presence of a reciprocal feedback loop among NAO, SIER, and SSTR. Subsequently, we move forward to demonstrate the affirmative nature of this loop. Emphasizing the skewness of NAO in Table (\ref{tab_skewness}), we offer insight into the bootstrap confidence intervals (C.I.) across different time intervals—daily, weekly, and monthly.  In a scenario of stable NAO, a skewness value of zero is expected. However, our findings unveil a negatively skewed distribution, indicating a statistically significant departure from stability. This pronounced outcome vividly underscores the prevailing instability within the NAO dynamics.

Together, these analyses substantiate the existence of a complex feedback loop among NAO, SIER, and SSTR. This discovery not only expands our understanding of climate interdependencies but also reveals a distinctive form of instability inherent within the North Atlantic system.

\subsection{Dynamic Statistical Approach}

\begin{figure}[h]
	\centering
	\includegraphics[width=0.3\linewidth]{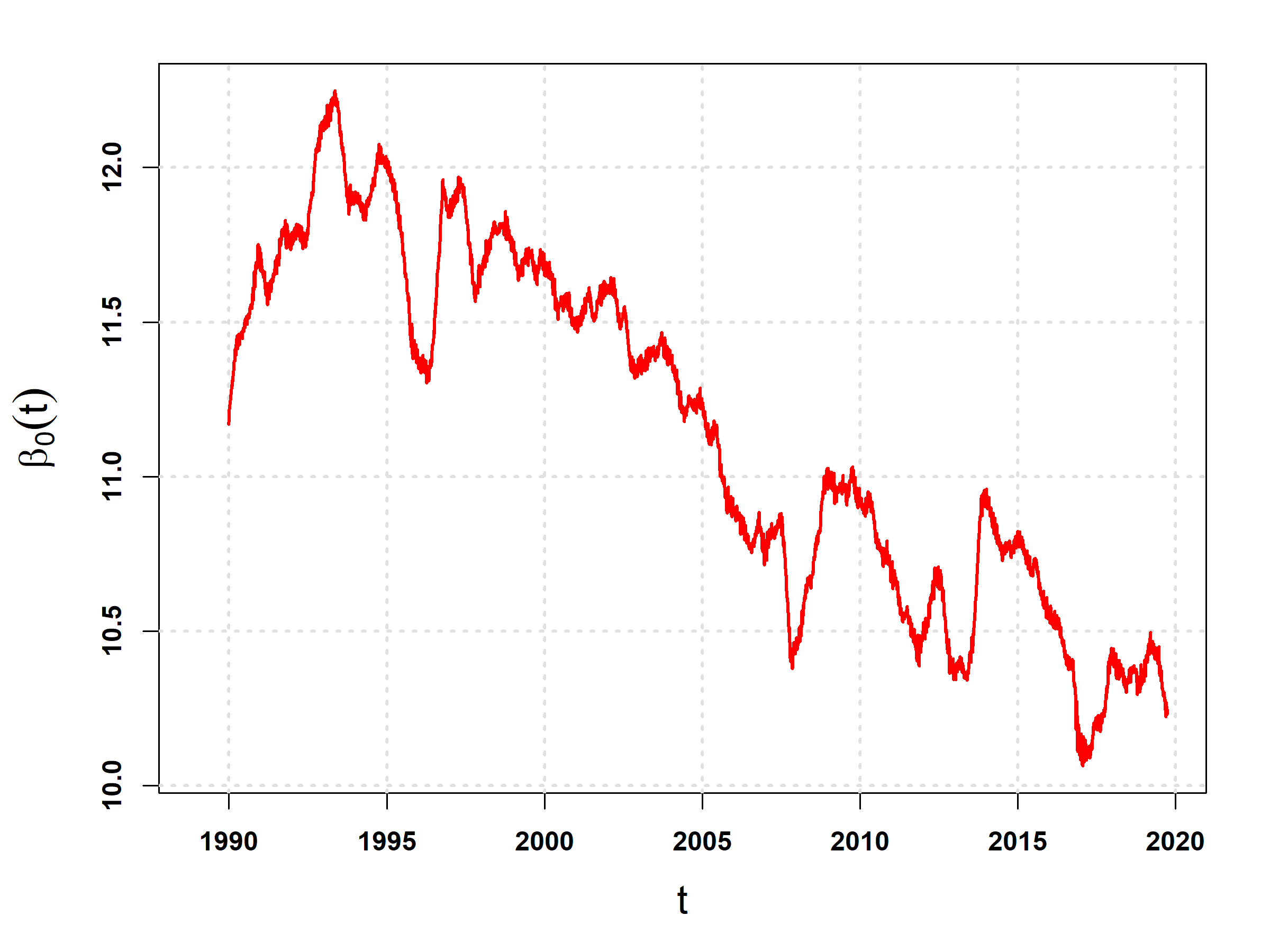}
        \llap{\parbox[b]{2in}{\textbf{(a)}\\\rule{0ex}{1.5in}}}
        \includegraphics[width=0.3\linewidth]{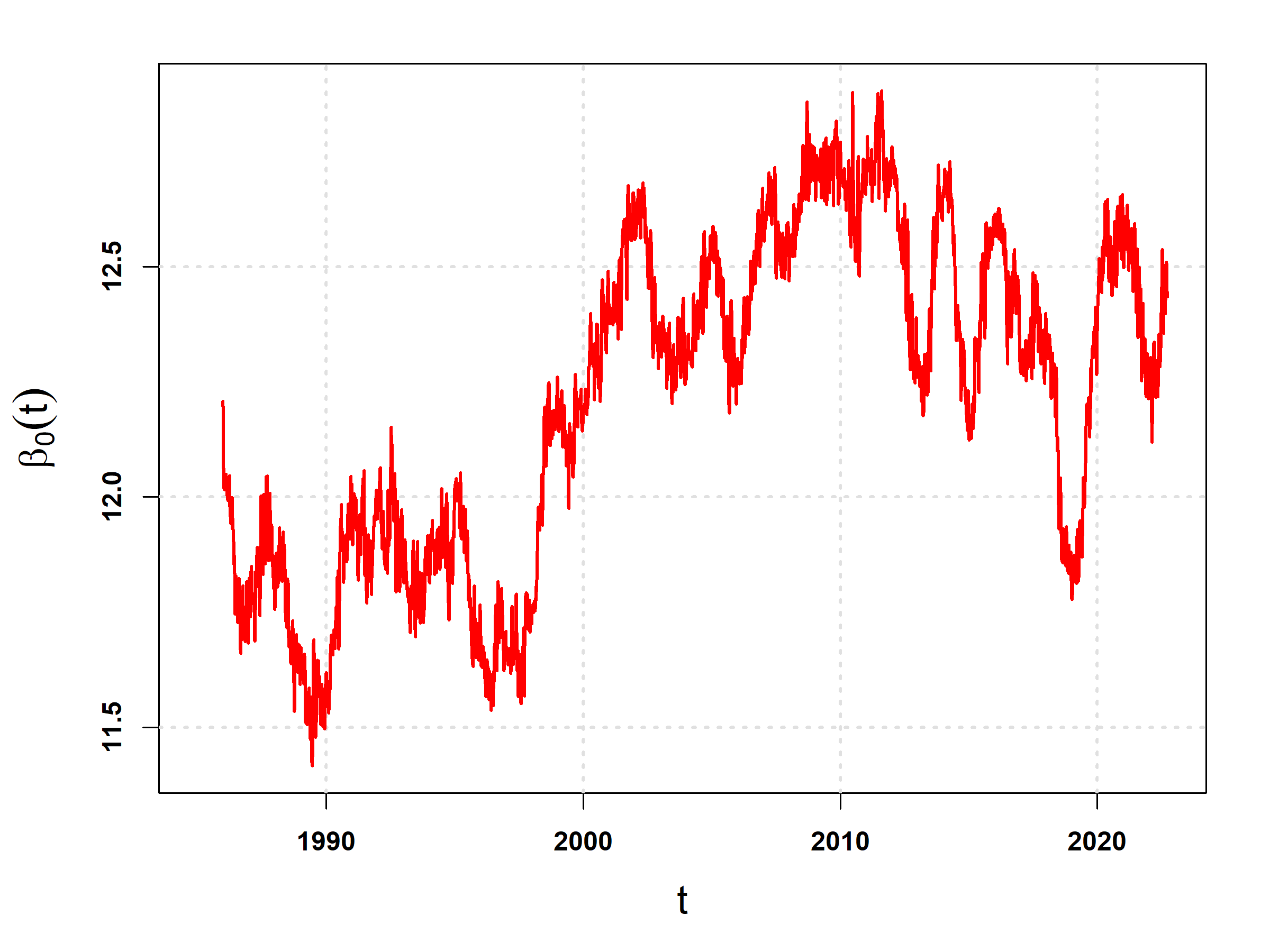}
        \llap{\parbox[b]{2in}{\textbf{(b)}\\\rule{0ex}{1.5in}}}
        \includegraphics[width=0.3\linewidth]{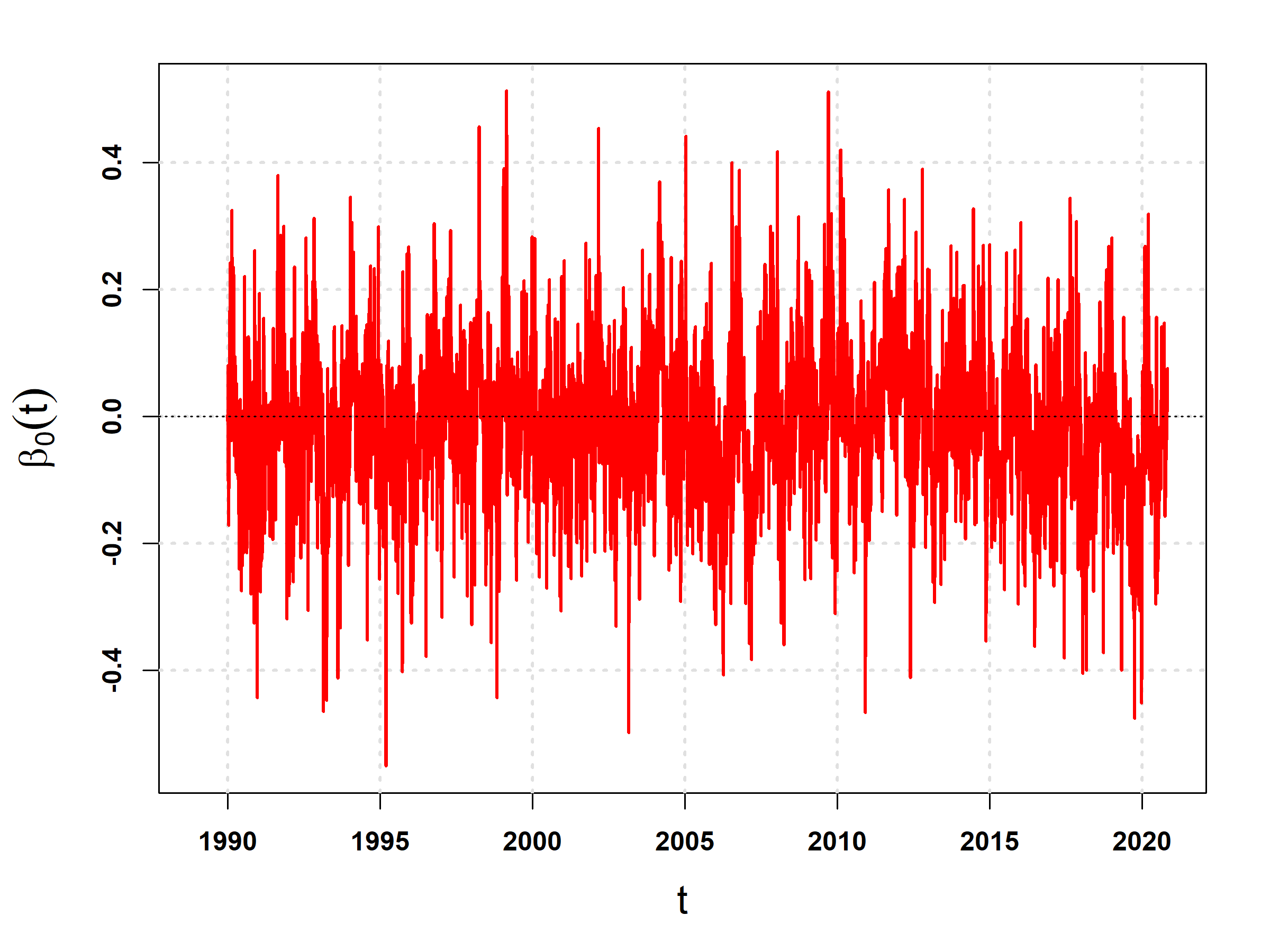}
        \llap{\parbox[b]{2in}{\textbf{(c)}\\\rule{0ex}{1.5in}}}
        \includegraphics[width=0.3\linewidth]{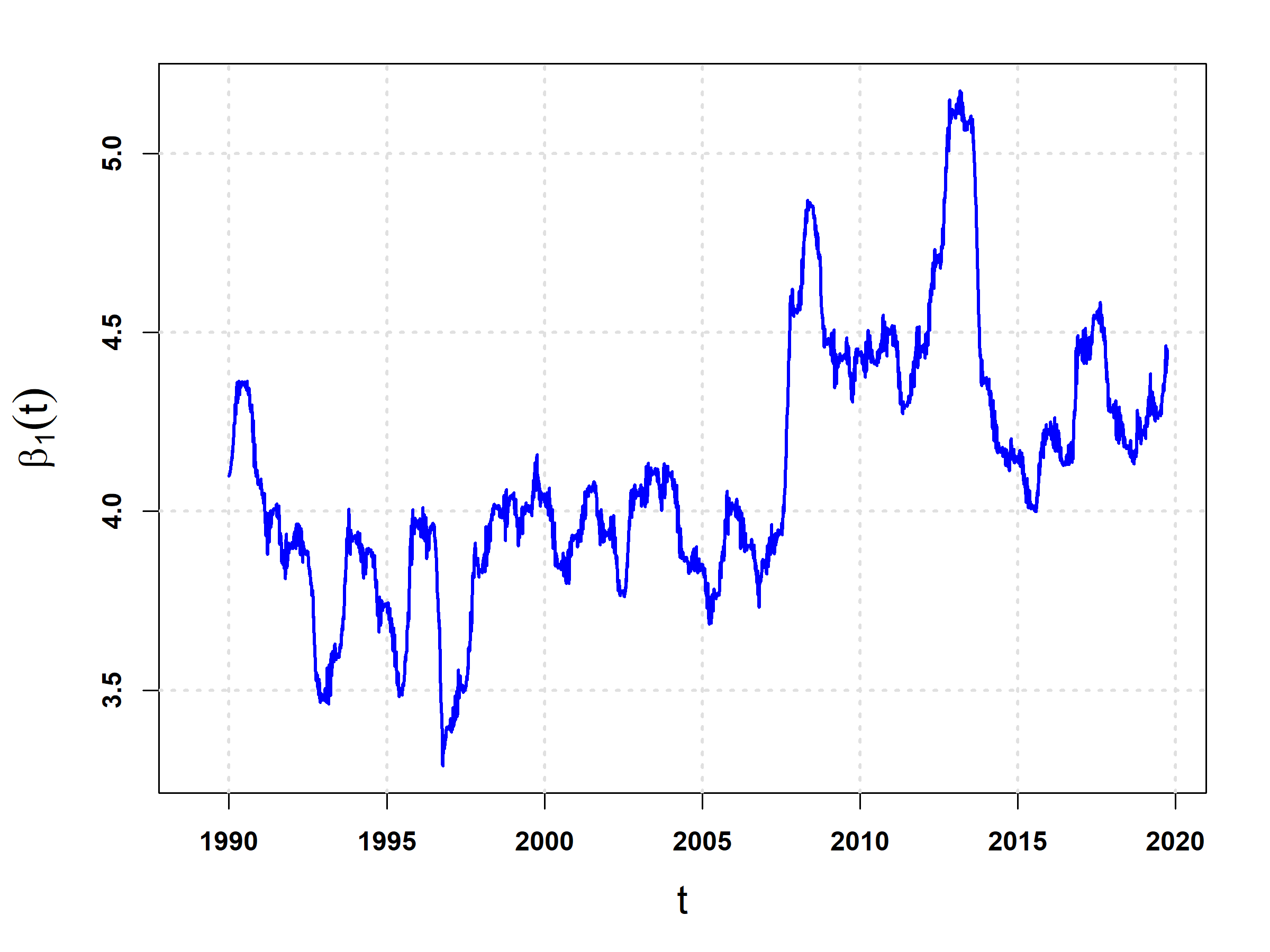}
        \llap{\parbox[b]{2in}{\textbf{(d)}\\\rule{0ex}{1.5in}}}
        \includegraphics[width=0.3\linewidth]{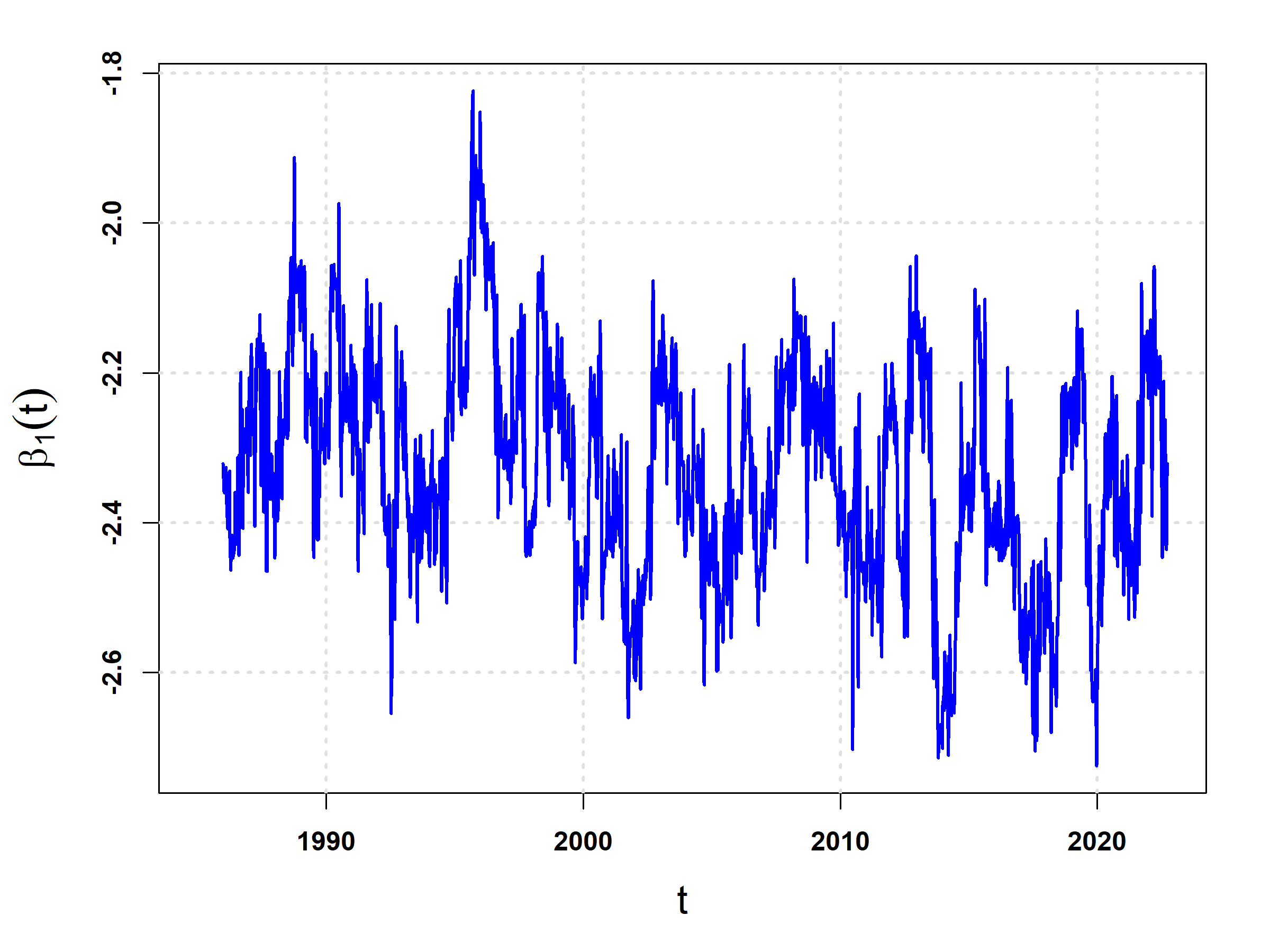}
        \llap{\parbox[b]{2in}{\textbf{(e)}\\\rule{0ex}{1.5in}}}
        \includegraphics[width=0.3\linewidth]{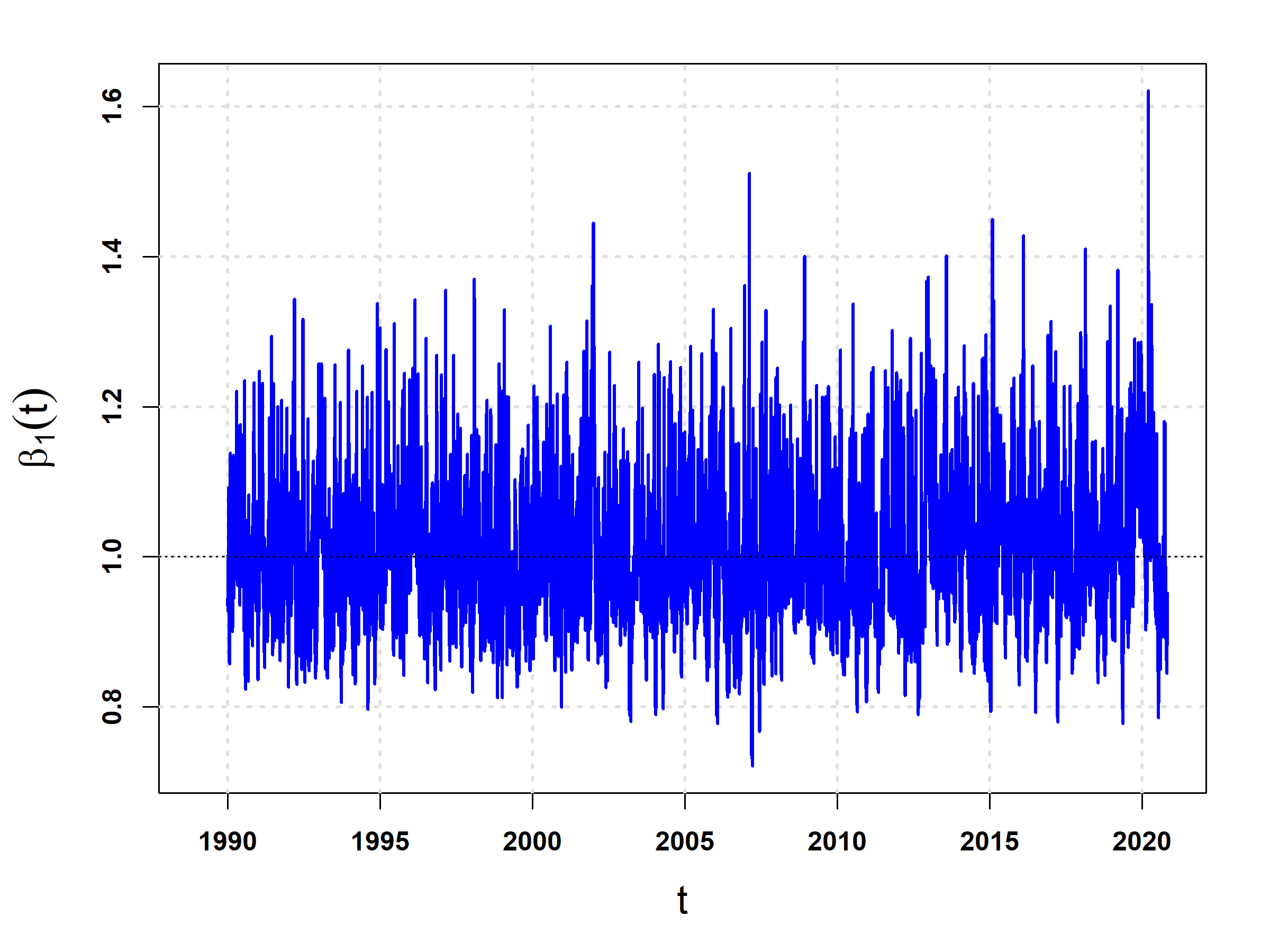}
        \llap{\parbox[b]{2in}{\textbf{(f)}\\\rule{0ex}{1.5in}}}
        \includegraphics[width=0.3\linewidth]{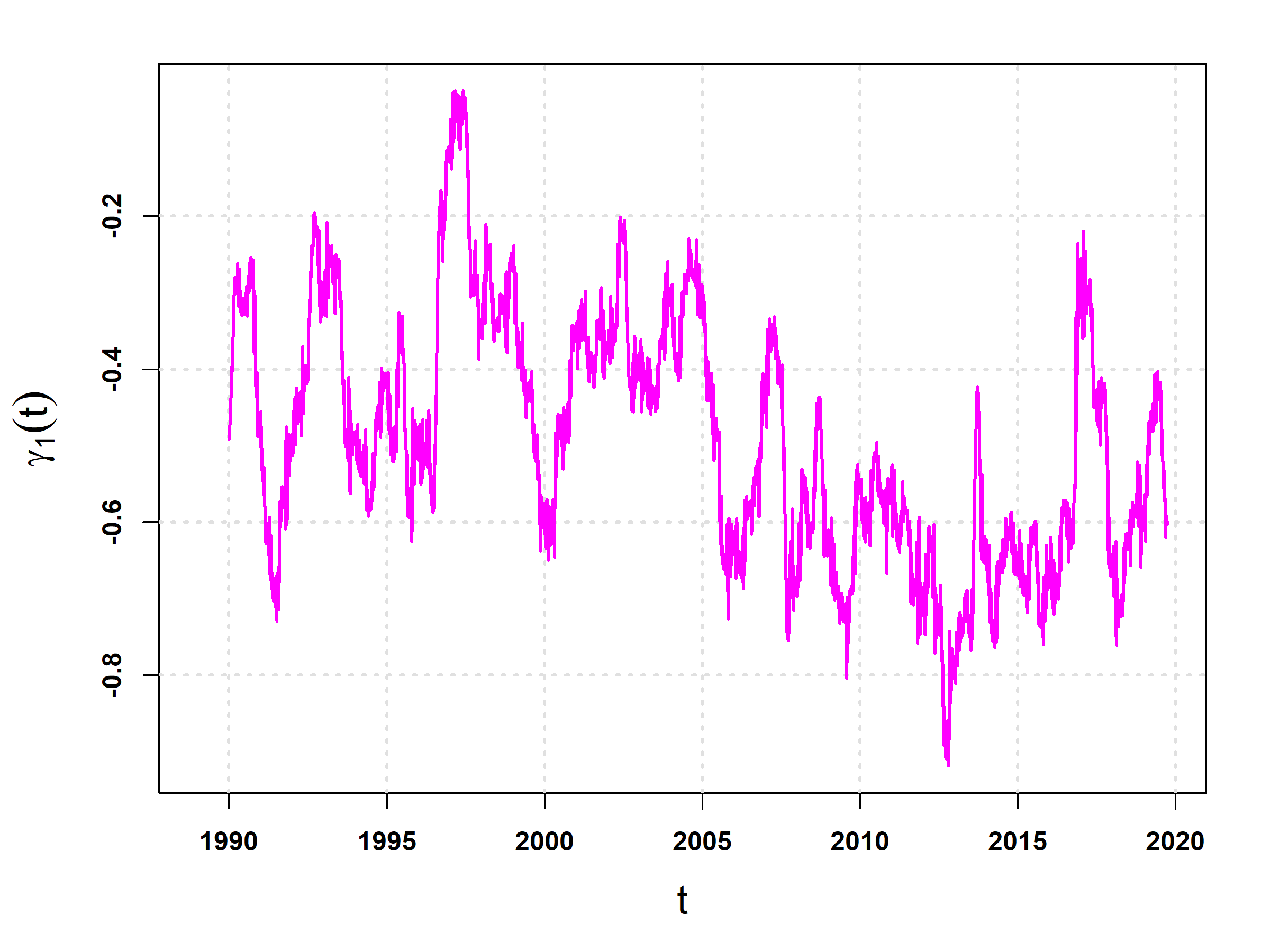}
        \llap{\parbox[b]{2in}{\textbf{(g)}\\\rule{0ex}{1.5in}}}
        \includegraphics[width=0.3\linewidth]{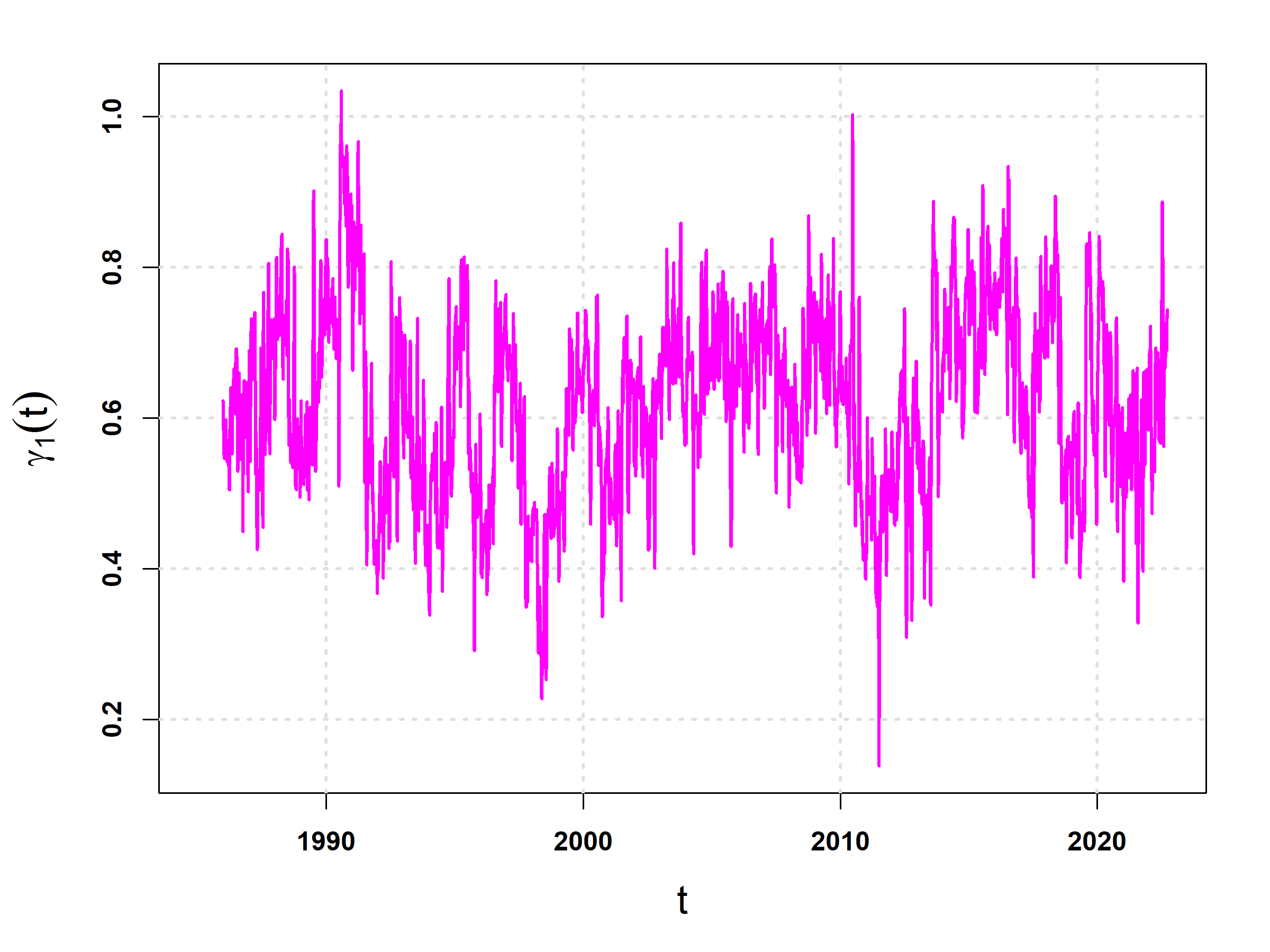}
        \llap{\parbox[b]{2in}{\textbf{(h)}\\\rule{0ex}{1.5in}}}
        \includegraphics[width=0.3\linewidth]{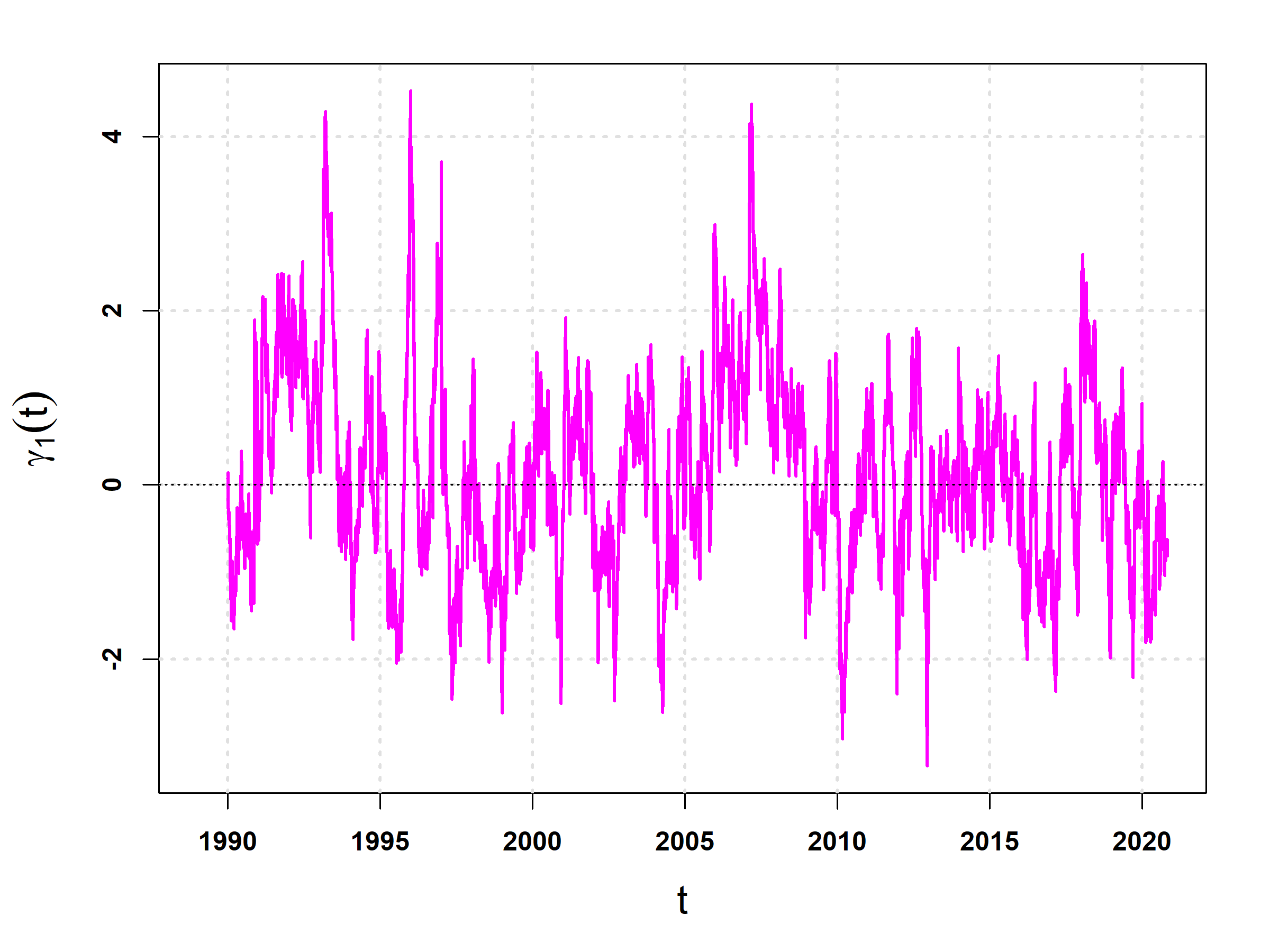}
        \llap{\parbox[b]{2in}{\textbf{(i)}\\\rule{0ex}{1.5in}}}
	\caption{(a) The downward trend in the dynamic intercept $\beta_0(t)$ of SIE indicates that the  SIE is shrinking over time.  (b) The downward trend in the dynamic intercept $\beta_0(t)$ of SST indicates an increasing trend in SST. (c) The dynamic intercept $\beta_0(t)$  of NAO is the mean-zero stationary process like NAO($t$). (d) The dynamic coefficient $\beta_1(t)$ corresponding to $\sin{\omega t}$ for SIE($t$). (e) The dynamic coefficient $\beta_1(t)$ corresponding to $\sin{\omega t}$ for SST($t$). (f) The dynamic coefficient $\beta_1(t)$ corresponding to $NAO(t-1)$ for $NAO(t)$. (g) The dynamic coefficient $\gamma_1(t)$ corresponding to $\cos{\omega t}$ for SIE($t$). (h) The dynamic coefficient $\gamma_1(t)$ corresponding to $\cos{\omega t}$ for SST($t$). (i) The dynamic coefficient $\gamma_1(t)$ corresponding to SST($t$) for NAO($t$).}
	\label{fig:beta_trend_SIE}
\end{figure}

In Figure (\ref{fig:beta_trend_SIE}a), the diminishing trajectory of the dynamic intercept $\beta_0(t)$ for SIE signifies a gradual decline in SIE over time. Correspondingly, Figure (\ref{fig:beta_trend_SIE}b) highlights the ascending trend of the dynamic intercept $\beta_0(t)$ for SST, indicating a progressive increase in SST. Transitioning to Figure (\ref{fig:beta_trend_SIE}c), the dynamic intercept $\beta_0(t)$ for NAO represents a mean-zero stationary process akin to NAO($t$).

Further exploration unveils Figure (\ref{fig:beta_trend_SIE}d), illustrating the dynamic coefficient $\beta_1(t)$ in harmony with $\sin{\omega t}$ for SIE($t$). Similarly, Figure (\ref{fig:beta_trend_SIE}e) elucidates the dynamic coefficient $\beta_1(t)$ corresponding to $\sin{\omega t}$ for SST($t$). In Figure (\ref{fig:beta_trend_SIE}f), the portrayal of the dynamic coefficient $\beta_1(t)$ pertains to $NAO(t-1)$ in relation to $NAO(t)$.

Additionally, Figures (\ref{fig:beta_trend_SIE}g), (\ref{fig:beta_trend_SIE}h), and (\ref{fig:beta_trend_SIE}i) provide insights into the dynamic coefficients $\gamma_1(t)$ associated with $\cos{\omega t}$ for SIE($t$), SST($t$), and NAO($t$), respectively.

Common criticisms of traditional statistical models often underscore the assumption of static relationships between predictors and dependent variables over time \citep{Kolstad2019}. To counter this limitation, our study adopts a BDLM as detailed in Section \ref{BDLM}, drawing inspiration from previous works i.e, \cite{Giovanni2009, Migon2010}. This innovative approach ensures an adaptive update of the predictor-dependent variable relationship as time unfolds (see Figure (\ref{fig:beta_trend_SIE})).

Furthermore, aligning  the findings of \cite{Kolstad2019}, our study substantiates a lagged correlation between NAO and SIE within the Barents-Kara Sea. Remarkably, our research extends beyond this region to encompass the broader North Atlantic and Arctic area, thereby broadening the scope of the observed relationship.

Overall, our methodology not only addresses the limitations of conventional statistical models but also enriches our understanding of the evolving relationships between key climate variables within the dynamic context of the North Atlantic region.

\section{Concluding Remarks}
\label{sec:conc}

In conclusion, this study delved into the intricate dynamics and identified critical instability driven by positive feedback loops among three pivotal climate variables: melting SIE, rising SST, and the NAO. Employing a generic approach rooted in statistical machine learning, we pursued a comprehensive analysis that offers valuable insights into climate variability and its implications for the North Atlantic region.

Unlike intricate climate forecast models, our methodology embraced a less computationally intensive yet more universal strategy, facilitating an all-encompassing examination across the vast North Atlantic area. Addressing a central critique highlighted by \cite{Kolstad2019}, our BDLM dynamically updated the predictor-dependent variable relationship, enabling us to overcome static model limitations.

The study's major revelations are noteworthy: (i) a reciprocal Granger causality between SIE and SST, (ii) a mutual Granger causality between SST and NAO, and (iii) an anti-correlation between SST and NAO. This anti-correlation implies that the increasing SST trend is likely to trigger increased occurrences of negative NAO. This aligns with our intriguing finding that the NAO index exhibits negative skewness at various time scales (daily, weekly, and monthly), contrary to its expected mean-zero stationary behavior.

Importantly, the negative skewness of the NAO index, despite its mean-zero stationary nature, signals an impending critical instability. This unsettling phenomenon suggests an elevated probability of negative NAO occurrences, foretelling increased bouts of frigid climates in the North Atlantic region, particularly affecting northern Europe and eastern North America. This realisation underscores the significance of this study in predicting a notable climate transformation.

Overall, this research contributes substantively to the understanding of critical instability within intricate climate systems. Leveraging techniques from statistical machine learning  and data science for complex systems \citep{chakrabarti2023data}, our study enhances our grasp of the dynamic interplay among vital climate variables, extending our insights into the intricate mechanisms that shape climate patterns.

\begin{table}[ht]
\begin{center}
	 \setlength{\arrayrulewidth}{.8mm}
	\begin{tabular}{p{8cm}p{2cm}p{2cm}}
	\hline
	\bf{Hurst Exponent}  & \bf{SIER} & \bf{SSTR}\\
		\hline
		Simple R/S Hurst estimation   & 0.77 & 0.81\\
		Corrected R over S Hurst exponent   & 0.83 &0.89\\
		Empirical Hurst exponent   &0.82 &  0.90\\
		Corrected empirical Hurst exponent  &0.81 & 0.89\\
		Theoretical Hurst exponent   & 0.53 & 0.52\\ \hline
	\end{tabular}
		 \caption{The Hurst Exponent value of the SIER, and SSTR using different methods.}
		 \label{tab_Hurst_exp_SIER_SSTR}
\end{center}
\end{table}

\begin{table}[ht] 
	\centering
\setlength{\arrayrulewidth}{.8mm}
\begin{tabular}{p{2cm}p{2cm}p{3cm}p{4cm}}
	\hline
		 & \bf{NAO} &\bf{ SIER} &  \bf{SSTR} \\
		\hline
		\bf{NAO } &  1.000 & 0.016 (0.063) &  -0.133 (\small{$<2.2*10^{-16}$})  \\
		\bf{SIER} & & 1.000 & -0.173 (\small{$<2.2*10^{-16}$})  \\
	\bf{SSTR } &    &  & 1.000 \\ 
	\hline
	\end{tabular}
		\caption{ Over a span of 38 years, from January 1982 to September 2019, the correlation matrix for NAO, SIER, and SSTR is examined. The accompanying P-values enclosed in parentheses provide insights into the significance of the correlations. Notably, the correlation between SSTR and NAO is statistically significant. Similarly, a robust correlation is observed between SIER and SSTR. However, the correlation between NAO and SIER exhibits relatively weak significance.}
	\label{tab_NAO_SST_SIE_corr}
\end{table}

\begin{table}[h]
\begin{center}
	 \setlength{\arrayrulewidth}{.8mm}
	\begin{tabular}{p{6cm}p{2.5cm}p{3cm}}
	\hline
	\bf{GC Models} &\bf{F-value} & \bf{p-value}\\
        \hline      
        \bf{ SSTR + SIER $\rightarrow$ NAO  } & 2.31 & 0.0178 \\  
        \bf{  NAO + SIER $\rightarrow$ SSTR } & 5.546 &$2.16\times10^{-6}$ \\ 
        \bf{   NAO + SSTR $\rightarrow$ SIER } & 7.714 & $2.27\times10^{-10}$\\ 
         \hline  
	\end{tabular}
		 \caption{ Table of F-value and p-value of different combinations of Granger Causal Models. Small p-values indicate that there is a feedback loop among NAO, SIER, and SSTR. }
		 \label{GC_models_result}
\end{center}
\end{table}

\begin{table}[ht]
\begin{center}
	 \setlength{\arrayrulewidth}{.8mm}
	\begin{tabular}{p{3cm}p{3cm}p{4cm}}
	\hline
	\bf{Period} & \bf{Skewness} & \bf{C.I.} \\
		\hline
		\textbf{Daily} &  -.210 & [-0.242, -0.169]\\
	    \textbf{Weekly}  & -.213 & [-0.305, -0.107]\\
		\textbf{Monthly}  &   -.194 & [-0.368, -0.005] \\ \hline
	\end{tabular}
		 \caption{The presented table highlights the skewness of NAO, along with bootstrap-derived confidence intervals (C.I.), across daily, weekly, and monthly time spans. While an anticipated stable NAO would exhibit a skewness of zero, our observations indicate a negatively skewed distribution. This statistically significant result underscores the presence of instability in NAO. }
		 \label{tab_skewness}
\end{center}
\end{table}

\section*{Data availability Statement}
The datasets used and/or analysed during the current study are available from the corresponding author on reasonable request.

\section*{Acknowledgements}
A.Y. is grateful for the fellowship from JNU and hospitality at CMI funded by AlgoLabs. S.D. acknowledges the partial financial support from Infosys Foundation, TATA Trust, and Bill \& Melinda Gates Foundation's grant to CMI.




\end{document}